\documentclass[a4paper,11pt]{article}
\pdfoutput=1 % if your are submitting a pdflatex (i.e. if you have
\usepackage{jheppub} % for details on the use of the package, please
\usepackage{natbib}
\usepackage{times}
\usepackage{float}
\usepackage{amsmath}
\usepackage{graphicx}
\usepackage{amssymb}
\usepackage{slashed}
\usepackage{xcolor}
\usepackage{soul} % For strikeout, using  \st{Hello world}
\usepackage{multirow}
\usepackage{hhline}
\usepackage{bbold}
\usepackage{booktabs}
\usepackage[T1]{fontenc} % if needed
\usepackage[utf8]{inputenc}
\usepackage{xfrac}

\usepackage{xspace}

\definecolor{violet}{cmyk}{0,1,0,0.2}

\allowdisplaybreaks

\makeatletter
\g@addto@macro\bfseries{\boldmath}
\makeatother

\title{Lepton-Quark Fusion at Hadron Colliders, precisely}

\author[a,b]{Admir Greljo,}
\author[c]{and Nud\v zeim Selimovi\'c}

\affiliation[a]{Albert Einstein Center for Fundamental Physics, Institut f\"{u}r Theoretische Physik, Universit\"{a}t Bern, \\ Sidlerstrasse 5, CH-3012 Bern, Switzerland.}
\affiliation[b]{CERN, Theoretical Physics Department, CH-1211 Geneva 23, Switzerland.}
\affiliation[c]{Physik-Institut, Universit\"{a}t Z\"{u}rich, CH-8057 Z\"{u}rich, Switzerland.}

\emailAdd{greljo@itp.unibe.ch}
\emailAdd{nudzeim@physik.uzh.ch}

\preprint{CERN-TH-2020-204, ZU-TH 49/20}

\abstract{
When a TeV-scale leptoquark has a sizeable Yukawa coupling, its dominant production mechanism at hadron colliders is the partonic-level lepton-quark fusion. Even though the parton distribution functions for leptons inside the proton are minuscule, they get compensated by the resonant enhancement. We present the first computation of higher order radiative corrections to the resonant leptoquark production cross section at the Large Hadron Collider (LHC). Next-to-leading (NLO) QCD and QED corrections are similar in size but come with the opposite sign. We compute NLO $K$-factors for a wide range of scalar leptoquark masses, as well as, all possible combinations of quark and lepton flavors and leptoquark charges. Theoretical uncertainties due to the renormalisation and factorisation scale variations and the limited knowledge of parton distribution functions are quantified. We finally discuss how to disentangle the flavor structure of leptoquark interactions by exploiting the interplay between different production channels.
}

\arxivnumber{2012.02092}

\begin{document}
	
\maketitle
	
\newpage

%%%%%%%%%%%%%%%%%%%%%%%%%%%%%%%%%%%%%%
	
\section{Introduction}
\label{intro}
%%%%%%%%%%%%%%%%%%%%%%%%%%%%%%%%%%%%%%

Leptoquarks (LQs) are hypothetical new bosons that convert quarks into leptons and vice versa. The discovery of a leptoquark would represent a major breakthrough in our understanding of particle interactions, pointing towards an underlying quark-lepton unification at short distances. The phenomenology of TeV-scale leptoquarks is a very rich and mature subject, for a recent review see Ref.~\cite{Dorsner:2016wpm}. Leptoquarks at the TeV-scale are consistent with the non-observation of proton decay and can be found in wildly different settings beyond the Standard Model (SM). For example, they are in the spectrum of low-scale quark-lepton unification models à la Pati-Salam (see e.g.~\cite{Pati:1974yy,DiLuzio:2017vat,Bordone:2017bld,Greljo:2018tuh,Fornal:2018dqn,Heeck:2018ntp,Cornella:2019hct,Blanke:2018sro,Balaji:2019kwe}). TeV-scale leptoquarks also appear as pseudo-Nambu–Goldstone bosons of a new strongly interacting dynamics possibly related to the origin of the electroweak symmetry breaking (see e.g.~\cite{Gripaios:2009dq,Fuentes-Martin:2020bnh,Barbieri:2017tuq,Sannino:2017utc,Marzocca:2018wcf}), or as a consequence of R-parity violation in supersymmetry (see e.g.~\cite{Giudice:1997wb,Csaki:2011ge,Altmannshofer:2020axr}). On the one hand, they lead to distinct indirect modifications of low-energy flavor transitions, neutrino properties, top quark, electroweak precision, and Higgs physics. On the other hand, the direct production of a leptoquark at the LHC leaves a remarkable signature in the detector. Namely, a leptoquark would appear as a resonance in the invariant mass of a lepton and a quark jet.

\begin{figure}[tbp]
\centering
\includegraphics[scale=0.7]{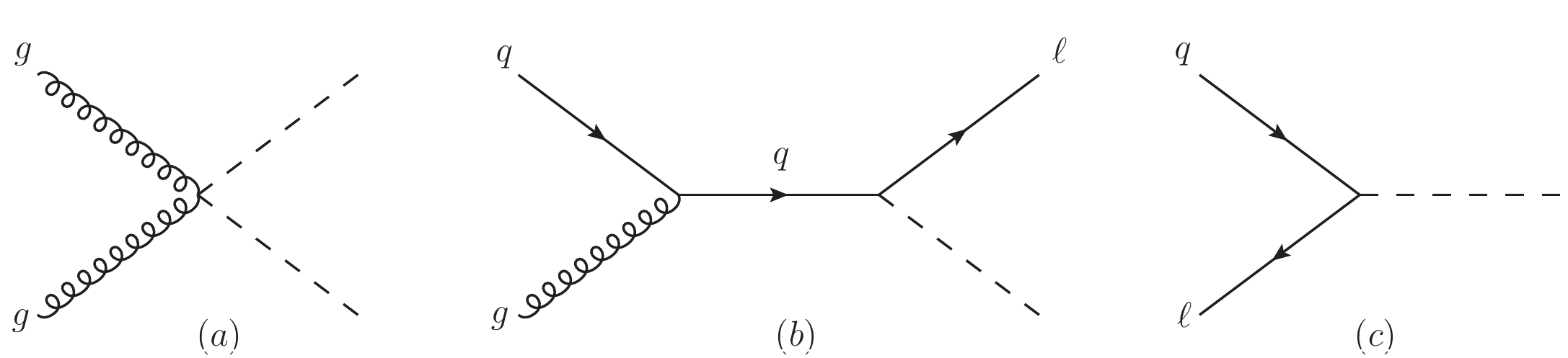}
\caption{\label{fig:diagram} Representative Feynman diagrams for three different leptoquark production mechanisms at hadron colliders: ($a$) pair $p p \to S_{Q_{\rm LQ}}^\dagger S_{Q_{\rm LQ}}$\,, ($b$) single $p p \to S_{Q_{\rm LQ}} \ell$ and ($c$) resonant $p p \to S_{Q_{\rm LQ}}$\,.}
\end{figure}

Leptoquarks are colored just like quarks. Therefore, they are copiously produced in pairs in proton-proton collisions at the LHC by strong force~\cite{Blumlein:1996qp,Kramer:1997hh,Kramer:2004df,Diaz:2017lit,Dorsner:2018ynv,Borschensky:2020hot,Allanach:2019zfr}. A representative Feynman diagram is shown in Fig.~\ref{fig:diagram} (a). In the limit of a small leptoquark coupling to quark and lepton ($y_{q\ell}$), the scalar leptoquark production at hadron colliders is determined entirely by the strong coupling $\alpha_s$ and the leptoquark mass $m_{\rm LQ}$. The phenomenology becomes more interesting once $y_{q\ell}$ is increased. This is particularly relevant when establishing a connection with the low-energy flavor physics. The present indirect constraints on a TeV-scale leptoquark suggest that $y_{q\ell}$ flavor matrix has a peculiar structure with some entries left unconstrained, and therefore possibly large. Taking a different perspective on the current data, in order to explain the existing experimental anomalies in $B$-meson decays~\cite{Lees:2013uzd, Hirose:2016wfn, Aaij:2015yra,
Aaij:2014ora, Aaij:2017vbb, Aaij:2013qta, Aaij:2015oid, Aaij:2019wad} or muon $g-2$~\cite{Bennett:2006fi}, some leptoquark couplings are required to be large. 

If leptoquarks are indeed behind the origin of these discrepancies, there will be other production mechanisms beyond the QCD-induced pair production. To begin with, for a sizeable $y_{q\ell}$, there is an additional contribution with $t$-channel lepton exchange in $q \bar q$ fusion~\cite{Bessaa:2014jya,Dorsner:2014axa}. However, the production of two leptoquarks becomes quickly phase-space suppressed with increasing leptoquark mass. Therefore, often discussed in the literature is the single leptoquark plus lepton production from quark-gluon scattering~\cite{Alves:2002tj,Hammett:2015sea,Mandal:2015vfa,Dorsner:2018ynv}. A representative Feynman diagram is shown in Fig.~\ref{fig:diagram} (b). The production cross section for this process is proportional to $|y_{q\ell} |^2$, but suffers less phase-space suppression. For a heavier leptoquark and a larger coupling, this production mechanism starts to dominate over the pair production. In this work, we are interested in a sizeable (yet perturbative) coupling range (i.e. $0.1 \lesssim y_{lq} \lesssim \sqrt{4 \pi}$ depending on the quark flavor), for which the production of a single leptoquark plus lepton becomes comparable or even favorable. 

\begin{figure}[tbp]
\centering
\includegraphics[scale=0.56]{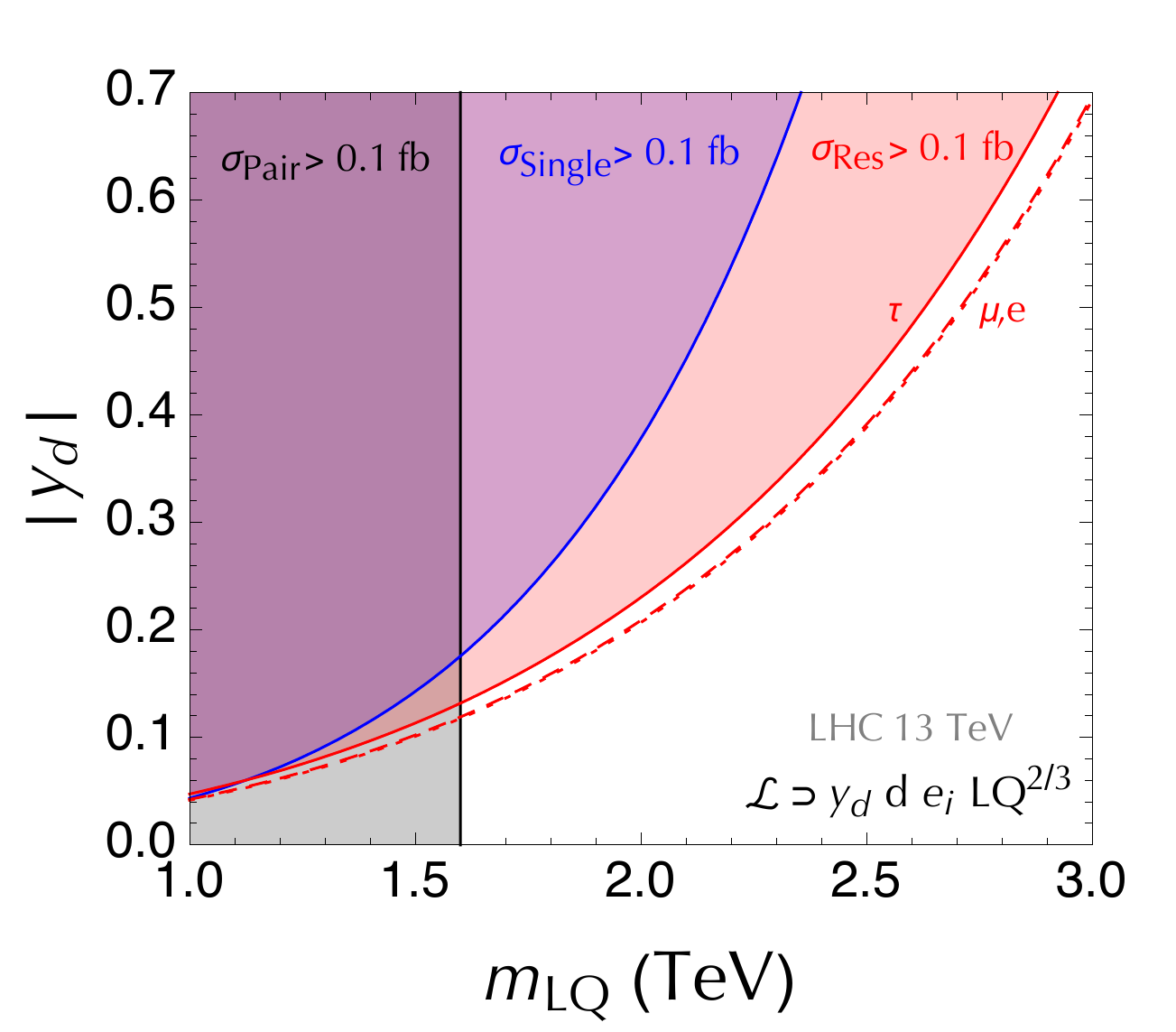}\;
\includegraphics[scale=0.56]{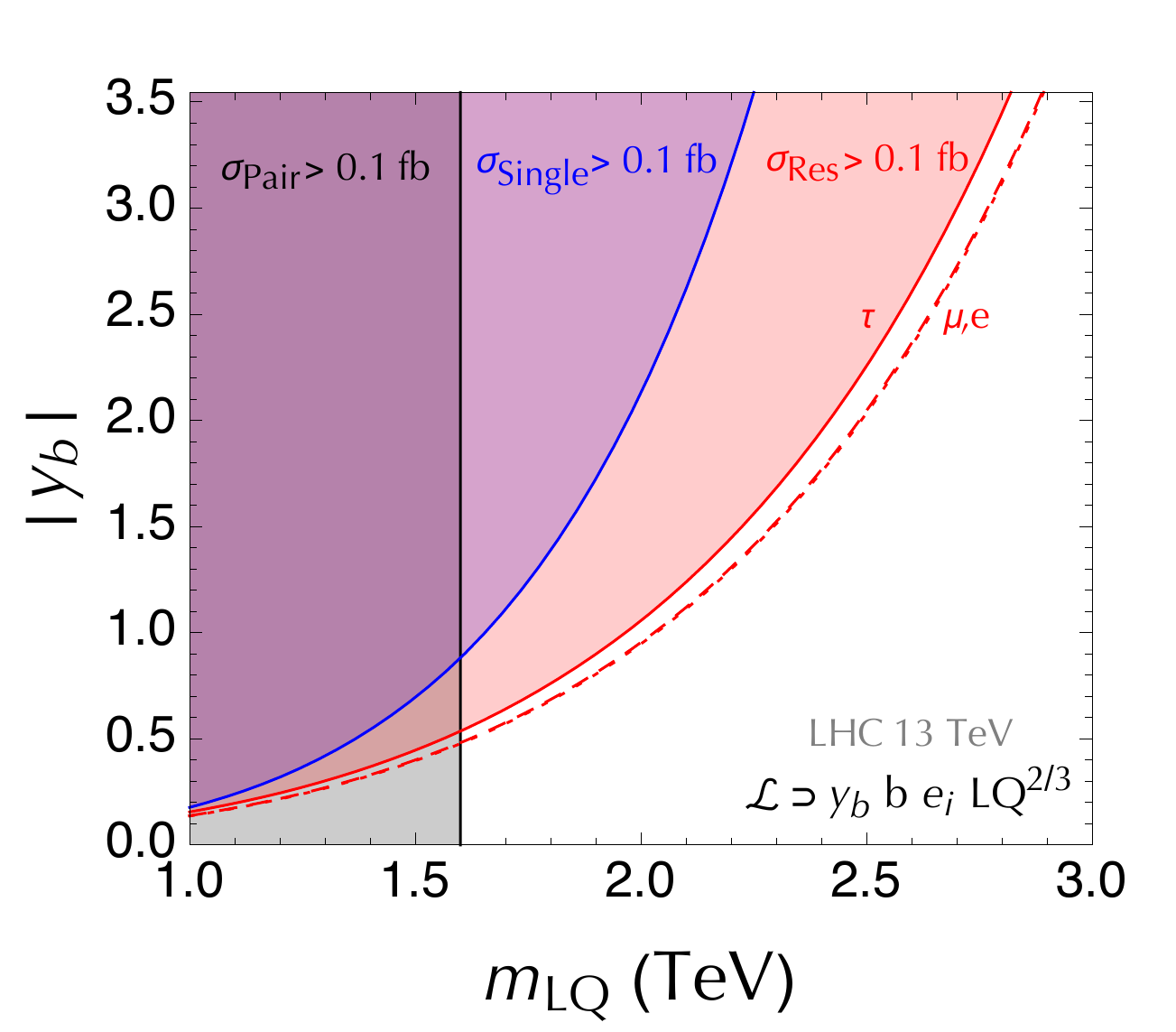}
\caption{\label{fig:xsecComparison} Comparison of cross sections for three leptoquark production mechanisms at the LHC ($\sqrt{s}=13$~TeV). Shaded regions show the parameter space in the leptoquark mass versus coupling plane for which the corresponding cross sections are $>0.1$~fb. The pair production $p p \to S_{Q_{\rm LQ}}^\dagger S_{Q_{\rm LQ}}$ cross section is shown in black, while the single $(p p \to S_{Q_{\rm LQ}}\, \ell)+$\,c.c. and the resonant $(p p \to S_{Q_{\rm LQ}}) +$\,c.c. cross sections are shown in blue and red, respectively. In the left (right) panel, the leptoquark interacts primarily with the down (bottom) quark. The lepton flavors in the resonant production are shown with solid ($\tau$), dashed ($\mu$) and dotted ($e$) lines. The electric charge of $S_{Q_{\rm LQ}}$ is set to $Q_{\rm LQ}=2/3$, however, the difference is negligible  for $Q=4/3$. For consistency, all cross sections are computed at NLO QCD (plus NLO QED for the resonant process) with the same central PDF set {\tt LUXlep-NNPDF31\_nlo\_as\_0118\_luxqed} (v2)~\cite{Buonocore:2020nai}. The first two processes are computed using the leptoquark toolbox~\cite{Dorsner:2018ynv}, while the resonant production is taken from Section~\ref{sec:dis}. The pair production from the $t$-channel leptoquark exchange is negligible in this coupling range. }
\end{figure}

For example, Fig.~\ref{fig:xsecComparison} shows the relative comparison of different channels in the mass versus coupling plane when the leptoquark couples to down quark (left panel) and bottom quark (right panel). The upper edge of the vertical axis is chosen such that the $t$-channel induced pair production is suppressed compared to the pure QCD contribution. Nonetheless, the single leptoquark production plus the charge-conjugated (c.c.) process, dominates over the pair production in the large portion of the parameter space shown in Fig.~\ref{fig:xsecComparison}.  Relevant information on these parameters can also be extracted from indirect leptoquark effects at high-$p_T$, such as Drell-Yan tails~\cite{Faroughy:2016osc,Schmaltz:2018nls,Greljo:2017vvb,Fuentes-Martin:2020lea,Greljo:2018tzh,Marzocca:2020ueu,Baker:2019sli}. These probe complementary parameter space compared to both single and pair production (see Section~4 in Ref.~\cite{Dorsner:2018ynv}).

The collider phenomenology of TeV-scale leptoquarks had a new twist recently. The precise extraction of lepton parton distribution functions (PDFs)~\cite{Buonocore:2020nai} based on the {\tt LUX} method~\cite{Manohar:2016nzj,Manohar:2017eqh} (see also~\cite{Bertone:2015lqa,Bertone:2017bme}) facilitated another leptoquark production mechanism, the {\it resonant leptoquark production}~\cite{Ohnemus:1994xf,Eboli:1997fb,Buonocore:2020erb}. The tree-level Feynman diagram is shown in Fig.~\ref{fig:diagram} (c). The production cross section for the direct lepton-quark fusion is also proportional to $|y_{q\ell} |^2$, but suffers even less phase-space suppression than the single leptoquark plus lepton channel. The difference between the two is the absence (presence) of a high-$p_T$ lepton. Therefore, the resonant channel cross section is always larger as shown in Fig.~\ref{fig:xsecComparison}.  Interestingly, this applies to all combinations of quarks and leptons involved. The ATLAS and CMS collaborations have extensively searched for leptoquarks in pair production and a single leptoquark plus lepton channel~\cite{Sirunyan:2017yrk,Sirunyan:2018jdk,Sirunyan:2018ryt,Aaboud:2019bye,Aad:2020jmj,ATLAS:2020sxq,Aad:2020iuy,Aad:2020sgw,Sirunyan:2018btu,Sirunyan:2018vhk,Sirunyan:2018xtm,Aaboud:2019jcc,CMS:2020gru}, however, the resonant production was not considered so far. Nonetheless, the phenomenological collider simulation in Ref.~\cite{Buonocore:2020erb} shows that the resonant channel has a potential to probe the uncharted territory of interest in the mass versus coupling plane. It is therefore of utmost importance for leptoquark hunters at the LHC to place the resonant production mechanism at the top of their to-do list.

In this paper we fill in the gap on the theory side. Leptoquark toolbox for precision collider studies~\cite{Dorsner:2018ynv} includes leptoquark pair and single production at NLO in QCD. The scope of this work is to precisely calculate the resonant leptoquark production cross at the LHC including for the first time higher order radiative corrections and quantify the uncertainties from the missing orders and limited knowledge of parton distribution functions. The main result of this paper are the resonant leptoquark production cross sections at the LHC at NLO QCD plus QED with the corresponding uncertainties. These are reported in Tables~\ref{tab:13up} and~\ref{tab:13down}, together with the complete set of NLO $K$-factors reported in Figs.~\ref{fig:SVelectron},~\ref{fig:SVmuon} and~\ref{fig:SVtau}. Interestingly, we find that NLO QED corrections are as important as QCD corrections. Along the way, we discuss the interplay between different production mechanism and propose methods to determine the quark flavor inside the proton from which the leptoquark was created. The present study is limited to scalar leptoquarks and will be extended to include vectors in the future. Radiative corrections in the scalar leptoquark models are not sensitive to the details of the ultraviolet completion, in contrast to the vector case~\cite{Fuentes-Martin:2019ign,Fuentes-Martin:2020luw,Fuentes-Martin:2020hvc}. 

The paper is organised as follows. In Section~\ref{sec:scalarLQ} we set up the framework and present compact analytic expressions for the relevant partonic cross sections stemming from loop calculations detailed in Appendices~\ref{sec:QEDNLO} and~\ref{sec:NLOQCD}. In Section~\ref{sec:dis} we perform a numerical calculation of the hadronic cross section for the resonant leptoquark production at the LHC using the most recent lepton parton distribution functions. Supplemental numerical results are left for Appendix~\ref{app:supplement}. We finally conclude in Section~\ref{sec:Conc}.

%%%%%%%%%%%%%%%%%%%%%%%%%%%%%%%%%%%%%%
\section{Scalar leptoquark resonant production}
\label{sec:scalarLQ}
%%%%%%%%%%%%%%%%%%%%%%%%%%%%%%%%%%%%%%

The inevitable condition for a field coupling quarks and leptons at the tree level is to transform in the (anti)fundamental representation of the $SU(3)$ part of the SM gauge group. The interaction between leptoquark and gluons is then completely specified and forms the basis for NLO QCD calculations. In contrast, the electroweak part of the SM allows for leptoquark representations involving different $SU(2)_L\times U(1)_Y$ multiplets with the corresponding hypercharges $Y$. As we are interested in evaluating the NLO QED corrections, the only relevant information is that after electroweak symmetry breaking, the possible absolute electric charges for any component of the $SU(2)_L$ multiplet are $|Q_{\rm LQ}|=\{1/3, 2/3, 4/3, 5/3\}$, in the units of the positron charge. Therefore, to assess the NLO QCD plus QED corrections to the resonant leptoquark production, we can treat one component inside the multiplet at a time, with the production of other components corresponding to separate processes. 

The fermion content is the SM one, and the quark-lepton interaction with the scalar leptoquark $S_{Q_{\rm LQ}}$ of charge $Q_{\rm LQ}$ is given by
\begin{equation}\label{eq:Lagdef}
    \mathcal{L} \supset - y^L_{q\ell} ~\bar q P_{L} \ell ~S_{Q_{\rm LQ}} - y^R_{q\ell} ~\bar q P_{R} \ell ~S_{Q_{\rm LQ}} ~+~{\rm h.c.}\,,
\end{equation}
where $y^{L,R}_{q\ell}$ are $3\times3$ matrices in flavor space, encoding the most general form of Yukawa couplings. The chiral fermionic fields $q_{L,R}$ and $\ell_{L,R}$ (note the left- and right-handed chiral projectors $P_{L,R}$) correspond to charge and mass eigenstates after the electroweak symmetry breaking. (Fermion mixings when going from the interaction to the mass basis, both in the quark and lepton sectors, are already absorbed in Yukawa matrices $y^{L,R}_{q\ell}$.) Depending on $Q_{\rm LQ}$, some fermionic fields in Eq.~\eqref{eq:Lagdef} are charge-conjugated from the usual SM definitions, for example $\mathcal{L} \supset - y^R_{ue} ~ \overline{u_L} \ e_L^C~S^\dagger_{1/3}$. We can use the proton composition to precipitate lepton-quark fusion involving quark flavors $u,d,s,c,b$ and charged leptons $e,\mu,\tau$. When calculating partonic cross sections, we will work in the limit of disregarding all fermion masses, which is an excellent approximation given the energy of the collisions. Neutrinos are not created in photon splitting and cannot be generated inside the proton at the order we are interested in. In the absence of fermion masses, possible interference terms involving left- and right-handed Yukawa couplings vanish. 
This allows us to independently treat processes in which leptoquark is resonantly produced by the same flavor combination of quarks and leptons, but of the opposite chirality. Additionally, the resonant leptoquark production is specified by one entry in the chiral Yukawa matrix irrespective of all other entries. When several flavor couplings contribute to the production of the same leptoquark, the individual contributions to production cross section factorise, and we add them separately. 

In full generality, we summarize that scalar leptoquarks (SLQs) are $SU(3)$ triplets, with four possible values of the electric charge, and their resonant production cross section is determined by the entries in the Yukawa matrices without interference. This exhausts all possibilities for SLQs and we conclude that our computation could be easily matched to any model containing these particles. Moreover, we note that neglecting fermion masses causes all one-loop corrections proportional to Yukawa couplings to vanish. Accordingly, for the case of SLQ, the dominant NLO effects originate from QED and QCD.\footnote{The situation is different in the case of vector leptoquarks (VLQs). The calculation of NLO corrections for these particles necessarily involves details depending on the UV completion that embeds them. For instance, in many popular extensions of the SM, VLQs are accompanied by the massive color octet affecting NLO QCD contributions to processes involving VLQ in a nontrivial way~\cite{Fuentes-Martin:2019ign,Fuentes-Martin:2020luw}. For the moment, we focus on the resonant production of the SLQs, while postponing a detailed analysis of spin-1 leptoquarks for future work.}

The relevant NLO QED (QCD) corrections to partonic cross section are calculated in Appendix~\ref{sec:QEDNLO} (\ref{sec:NLOQCD}). The hadronic cross section is obtained after convoluting the relevant partonic cross sections, $\hat{\sigma}$, with the parton distribution functions, $f_i$ and $f_j$, in the following way,
\begin{equation}\label{eq:xsecsve}
\sigma(s) = 2\ \sum_{ij}  \int_{\xi}^{1}dy \, f_i(y)\int_{\xi/y}^1 dz\, \frac{\xi}{y z^2} \ f_j\left( \frac{\xi}{yz}\right)\ \hat{\sigma}_{ij}(z)~,
\end{equation}
where $\xi = m^2_{{\rm LQ}}/s$, $\sqrt{s}$ is the collider center of mass energy, {$y$ is the fraction of proton momentum carried by the parton labeled by $i$, and $z=m_{\rm LQ}^2/\hat{s}$, with $\hat{s}$ being the partonic-level center of mass energy.} The sum goes over $ij = \{q\ell$, $g\ell$, $q \gamma$\}, with the individual cases corresponding to Eqs.~\eqref{eq:xsecql}, ~\eqref{eq:xsecgl}, and ~\eqref{eq:xsecqgamma}, respectively.

%%%%%%%
\subsection{Next-to-leading order QCD corrections}
%%%%%%%

The hadronic cross section for resonant leptoquark production is set by the size of the colliding parton densities, and the size of the parton level cross section. The Yukawa couplings are $\mathcal{O}
(1)$, and at the leading order (LO), the partonic cross section scales as $\hat{\sigma}_0\propto |y_{ql}|^2$.  The parton density for gluons and quarks can be viewed as a sum of terms $\sum_n (\alpha_s L)^n$, where $\alpha_s$ is the QCD coupling, $L=\log(\mu_{F}^2/\Lambda^2)$, with $\mu_F$ representing the factorisation scale, and $\Lambda$ is the typical hadronic scale. The QCD coupling is evaluated at the factorisation scale and its size is set by $\alpha_s\approx 1/L$. We conclude that gluon and quark PDFs are non-perturbative objects of $\mathcal{O}(1)$. In contrast, the photon density is the first order QED effect and its size is determined by $\alpha L \sum_n (\alpha_s L)^n$, where $\alpha$ is the QED coupling. Further, as a result of photon splitting, lepton PDFs are generated at the next order in QED and their size is given by $\alpha^2 L^2 \sum_n (\alpha_s L)^n$. We apply the same QED to QCD coupling comparison already employed in~\cite{Buonocore:2020nai,Manohar:2016nzj,Manohar:2017eqh} and use that $\alpha\approx\alpha_s^2$. Accordingly, in terms of $\alpha_s$, the size of the photon density is $\mathcal{O}(\alpha_s)$, while the lepton densities are $\mathcal{O}(\alpha_s^2)$. The size of the LO hadronic cross section for resonant leptoquark production is then $\int (f_q \otimes f_{\ell}) \ \hat{\sigma}_0 \sim \mathcal{O}(\alpha_s^2)$. Therefore, the typical QCD correction coming from $\mathcal{O}(\alpha_s)$ diagrams represents the contribution to hadronic cross section which is $\mathcal{O}(\alpha_s^3)$.

The virtual corrections from gluon loops (Fig.~\ref{fig:QCDvirtual}) are summed with the diagrams involving the real gluon emission (Fig.~\ref{fig:QCDquark}) to obtain the IR safe partonic cross section
	\begin{gather}
	\label{eq:xsecql}
    \hat{\sigma}_{q\ell}(z) = \frac{\pi z|y_{q\ell}|^2}{4m_{\rm LQ}^2}\Bigg\{\left[1+\frac{\alpha_s}{2\pi}C_F\left(\frac{3}{2}\log\left(\frac{\mu_{\rm R}^2}{\mu_{\rm F}^2}\right)-\frac{\pi^2}{3}\right)\right]\delta(1-z)-\frac{2z}{(1-z)_{+}}\\\nonumber
    +2(1+z^2)\left(\frac{\log(1-z)}{(1-z)}\right)_{+}-\frac{1+z^2}{(1-z)_{+}}\log\left(\frac{z\mu_{\rm F}^2}{m_{\rm LQ}^2}\right)\Bigg\}\,,
    \end{gather}
where $C_F=4/3$ and $\mu_{\rm F}$, $\mu_{\rm R}$ are the factorisation and renormalisation scales, respectively. The remaining $\mathcal{O}(\alpha_s)$ diagrams that contribute to the resonant leptoquark production involve gluons in the initial state (Fig.~\ref{fig:QCDglue}). The partonic cross section in this case reads
	\begin{equation}
\hat{\sigma}_{g\ell}(z) = \frac{\pi z|y_{q\ell}|^2}{4m_{\rm LQ}^2}\frac{\alpha_s}{2\pi}T_R\left[-\log\left(\frac{z\mu^2_{\rm F}}{(1-z)^2m_{\rm LQ}^2}\right)(z^2+(1-z)^2)+2z(1-z)(2+\log z)\right],
\label{eq:xsecgl}
\end{equation}
	where $T_R=1/2$. The $\overline{\rm MS}$ scheme was utilized both for factorisation and renormalisation. The NLO QCD corrections are universal for all leptoquark types. More details about the partonic cross section calculation can be found in Appendices \ref{sec:QEDNLO} and \ref{sec:NLOQCD}.

%%%%%%%
\subsection{Next-to-leading order QED corrections}
\label{sec:QEDmain}
%%%%%%%
 The NLO QED corrections are provided by processes where the initial lepton is replaced by a photon splitting into lepton pairs (Fig.~\ref{fig:QEDdown}). We estimate the size of these corrections by $\alpha_s$ power counting for the leptoquark production via $\gamma+q \to \ell +$\,LQ. When convoluted with the corresponding PDFs, the size of the resonant cross section is
\begin{equation}
\int \left(f_q \otimes f_\gamma \right) \hat{\sigma}_{\gamma q} \approx 1\times\alpha L\times  \alpha = \alpha_s^3~.
\end{equation}
Interestingly enough, the QED corrections are of the same order as the typical QCD corrections and their inclusion is essential in assessing the NLO effects in resonant leptoquark production. Employing the $\overline{\rm MS}$ factorisation scheme, the partonic cross section reads
\begin{equation}
\hat{\sigma}_{q\gamma} = \frac{\pi z|y_{q\ell}|^2}{4m_{\rm LQ}^2}\frac{\alpha}{2\pi} \left(-\log\left(\frac{z\ \mu_{\rm F}^2}{(1-z)^2 m_{\rm LQ}^2}\right)(z^2+(1-z)^2)+X_{Q_{\rm LQ}}(z)\right).
\label{eq:xsecqgamma}
\end{equation}
The logarithmic part is universal for all leptoquark types since it originates from photon splitting to charged lepton pair, while the charge dependence is encoded in the functions $X_{Q_{\rm LQ}}(z)$,
\begin{align}
X_{1/3}(z) &=  - \frac{2}{9}(1-z)(5-13z)
+ \frac{2}{9}(1-5z)z\log z,\\
X_{2/3}(z) &= - \frac{11}{18}(1-z)(1-5z)
+ \frac{8}{9}(1-2z)z\log z,\\
X_{4/3}(z) &=  \frac{1}{18}(1-z)(13+103z)
+ \frac{16}{9}(2-z)z\log z,\\
X_{5/3}(z) &=   \frac{2}{9}(1-z)(7+37z)
+ \frac{10}{9}(5-z)z\log z,
\end{align}
where subscripts $\{1/3,\ 2/3,\ 4/3,\ 5/3\}$ correspond to electric charge of the leptoquark. Since the CP is conserved, the same formulas hold for the charge-conjugated processes. Note that loop diagrams involving photons are higher order in QCD coupling. Namely, the size of these diagrams, for the 1-loop corrections to partonic-level cross section involving photons, $\hat{\sigma}_{q\ell}^{(1)}$, is given by
\begin{equation}
\int \left(f_q \otimes f_\ell \right) \hat{\sigma}_{q\ell}^{(1)} \approx 1\times\alpha^2 L^2\times  \alpha = \alpha_s^4~,
\end{equation}
and we neglect them. The detailed derivation of the NLO QED corrections is presented in Appendix~\ref{sec:QEDNLO}.
	
%%%%%%%%%%%%%%%%%%%%%%%%%%%%%%%%%%%%%%
\section{Numerical results and discussion}
\label{sec:dis}
%%%%%%%%%%%%%%%%%%%%%%%%%%%%%%%%%%%%%%

We carry out a numerical calculation of the hadronic cross section for the resonant leptoquark production in $pp$ collisions. We consider the most general flavor structure of the leptoquark coupling $y_{q\ell}$ to a quark $q \equiv d_L, d_R$, $u_L, u_R$, $s_L, s_R$, $c_L, c_R, b_L$ or $b_R$, and a lepton $\ell \equiv e_L, e_R$, $\mu_L, \mu_R, \tau_L$, or $\tau_R$. All options for the leptoquark electric charge are considered, $|Q_{{\rm LQ}}|=1/3$, $2/3$, $4/3$ and $5/3$. Cross sections are calculated for every $q\ell$ combination separately assuming $y_{q \ell} = 1$. As a reminder, the total cross section is simply the sum over different channels, $\sigma = \sum_{q,\ell} |y_{q \ell}|^2 \sigma_{q \ell}$. We compute the process and its charge conjugate at leading and next-to-leading order in QCD and QED. We scan over the large leptoquark mass window $m_{{\rm LQ}}=[500 - 5000]$~GeV relevant for the future studies at the LHC. As a benchmark, the collider center of mass energy is set to $\sqrt{s} = 13$~TeV.

\begin{table}
\centering
		{\small
			\begin{tabular}{|c|c|c|c|}
				\hline
				$m_\mathrm{LQ}$\,[TeV]& Partons &$\sigma_\mathrm{S^{1/3}}$ [pb]&$\sigma_\mathrm{S^{5/3}}$ [pb]\\
				\hline
				\hline
				\multirow{6}{*}{0.9} & $u\ +\ e$ & $(1.45\times 10^{-1})^{+3.1\%}_{-3.7\%}\pm 1.8\%$ & $(1.58\times 10^{-1})^{+2.9\%}_{-3.4\%}\pm 1.8\%$ \\\cline{2-4}
				
				 &$u\ +\ \mu$ & $(1.39\times 10^{-1})^{+3.1\%}_{-3.8\%}\pm 1.9\%$ & $(1.52\times 10^{-1})^{+2.9\%}_{-3.5\%}\pm 1.8\%$ \\\cline{2-4}
				 
				 &$u\ +\ \tau$ & $(1.11\times 10^{-1})^{+3.6\%}_{-4.0\%}\pm 2.0\%$ & $(1.23\times 10^{-1})^{+3.4\%}_{-3.6\%}\pm 2.0\%$ \\
				 \hhline{|~|===|}
				 
				 &$c\ +\ e$& $(1.32\times 10^{-2})^{+4.2\%}_{-5.1\%}\pm 12.1\%$ & $(1.44\times 10^{-2})^{+3.9\%}_{-4.7\%}\pm 12.2\%$ \\\cline{2-4}
				 
				 &$c\ +\ {\mu}$ & $(1.29\times 10^{-2})^{+4.3\%}_{-5.2\%}\pm 12.0\%$ & $(1.40\times 10^{-2})^{+3.9\%}_{-4.8\%}\pm 12.0\%$ \\ \cline{2-4}
				 
				 &$c\ +\ {\tau}$ & $(1.01\times 10^{-2})^{+4.6\%}_{-5.5\%}\pm 12.2\%$ & $(1.12\times 10^{-2})^{+4.1\%}_{-5.0\%}\pm 12.2\%$ \\

				\hline
				\hline
				\multirow{6}{*}{1.6} & $u\ +\ {e}$ & $(1.40\times 10^{-2})^{+2.8\%}_{-3.3\%}\pm 2.0\%$ & $(1.49\times 10^{-2})^{+2.7\%}_{-3.1\%}\pm 2.0\%$ \\\cline{2-4}
				
				 &$u\ +\ {\mu}$ & $(1.36\times 10^{-2})^{+2.9\%}_{-3.4\%}\pm 2.0\%$ & $(1.46\times 10^{-2})^{+2.7\%}_{-3.1\%}\pm 2.0\%$ \\ \cline{2-4}
				 
				 &$u\ +\ {\tau}$ & $(1.11\times 10^{-2})^{+3.3\%}_{-3.5\%}\pm 2.2\%$ & $(1.20\times 10^{-2})^{+2.9\%}_{-3.2\%}\pm 2.2\%$ \\ \hhline{|~|===|}
				 
				 &$c\ +\ {e}$& $(7.31\times 10^{-4})^{+3.8\%}_{-4.4\%}\pm 24.2\%$ & $(7.80\times 10^{-4})^{+3.6\%}_{-4.1\%}\pm 24.3\%$ \\ \cline{2-4}
				 
				 &$c\ +\ {\mu}$ & $(7.16\times 10^{-4})^{+3.8\%}_{-4.4\%}\pm 24.0\%$ & $(7.65\times 10^{-4})^{+3.6\%}_{-4.1\%}\pm 24.1\%$ \\ \cline{2-4}
				 
				 &$c\ +\ {\tau}$ & $(5.78\times 10^{-4})^{+4.1\%}_{-4.7\%}\pm 24.2\%$ & $(6.28\times 10^{-4})^{+3.8\%}_{-4.3\%}\pm 24.3\%$ \\
				 
				\hline
				\hline
				\multirow{6}{*}{2.5} & $u\ +\ {e}$ & $(1.53\times 10^{-3})^{+2.6\%}_{-3.0\%}\pm 2.4\%$ & $(1.61\times 10^{-3})^{+2.5\%}_{-2.8\%}\pm 2.4\%$ \\\cline{2-4}
				
				 &$u\ +\ {\mu}$ & $(1.50\times 10^{-3})^{+2.6\%}_{-3.0\%}\pm 2.4\%$ & $(1.59\times 10^{-3})^{+2.5\%}_{-2.9\%}\pm 2.4\%$ \\ \cline{2-4}
				 
				 &$u\ +\ {\tau}$ & $(1.25\times 10^{-3})^{+2.9\%}_{-3.2\%}\pm 2.5\%$ & $(1.33\times 10^{-3})^{+2.6\%}_{-3.0\%}\pm 2.5\%$ \\ \hhline{|~|===|}
				 
				 &$c\ +\ {e}$& $(5.52\times 10^{-5})^{+3.3\%}_{-3.8\%}\pm 41.9\%$ & $(5.83\times 10^{-5})^{+3.1\%}_{-3.6\%}\pm 42.1\%$ \\ \cline{2-4}
				 
				 &$c\ +\ {\mu}$ & $(5.43\times 10^{-5})^{+3.3\%}_{-3.8\%}\pm 41.8\%$ & $(5.74\times 10^{-5})^{+3.2\%}_{-3.6\%}\pm 41.9\%$ \\ \cline{2-4}
				 
				 &$c\ +\ {\tau}$ & $(4.48\times 10^{-5})^{+3.5\%}_{-4.0\%}\pm 42.0\%$ & $(4.79\times 10^{-5})^{+3.3\%}_{-3.8\%}\pm 42.1\%$ \\
				 
				\hline
				\hline
				\multirow{6}{*}{4.0} & $u\ +\ {e}$ & $(7.21\times 10^{-5})^{+2.3\%}_{-2.6\%}\pm 3.1\%$ & $(7.49\times 10^{-5})^{+2.2\%}_{-2.5\%}\pm 3.0\%$ \\\cline{2-4}
				
				 &$u\ +\ {\mu}$ & $(7.14\times 10^{-5})^{+2.3\%}_{-2.6\%}\pm 3.1\%$ & $(7.42\times 10^{-5})^{+2.2\%}_{-2.5\%}\pm 3.1\%$ \\\cline{2-4}
				 
				 &$u\ +\ {\tau}$ & $(6.02\times 10^{-5})^{+2.4\%}_{-2.7\%}\pm 3.2\%$ & $(6.30\times 10^{-5})^{+2.3\%}_{-2.6\%}\pm 3.2\%$ \\\hhline{|~|===|}
				 
				 &$c\ +\ {e}$& $(2.35\times 10^{-6})^{+2.6\%}_{-3.0\%}\pm 63.0\%$ & $(2.45\times 10^{-6})^{+2.6\%}_{-2.9\%}\pm 63.1\%$ \\\cline{2-4}
				 
				 &$c\ +\ {\mu}$ & $(2.33\times 10^{-6})^{+2.7\%}_{-3.0\%}\pm 62.9\%$ & $(2.42\times 10^{-6})^{+2.6\%}_{-2.9\%}\pm 63.1\%$ \\\cline{2-4}
				 
				 &$c\ +\ {\tau}$ & $(1.96\times 10^{-6})^{+2.8\%}_{-3.1\%}\pm 63.1\%$ & $(2.05\times 10^{-6})^{+2.7\%}_{-3.0\%}\pm 63.2\%$\\
				 
				\hline
\end{tabular}}
\caption{\label{tab:13up} Inclusive cross sections in {\bf pb} for the resonant leptoquark production from up-type quarks, $p p \to $\,LQ + charge-conjugated process, as a function of the leptoquark mass $m_{{\rm LQ}}$ at $\sqrt{s}=13$\,TeV. The cross section $\sigma_\mathrm{S^{1/3}}\; (\sigma_\mathrm{S^{5/3}})$ corresponds to the resonant production of scalar LQ with absolute electric charge $1/3\ (5/3)$ when the associated Yukawa coupling strength is set to one, $y_{q\ell} = 1$. The second column denotes which quark-lepton pair couples to the corresponding leptoquark. First (second) uncertainty is due to the renormalisation and factorisation scale variations (PDF replicas),  and is given in per cent units. See Section~\ref{sec:dis} for details.}
\end{table}

\begin{table}
\centering
		{\small
			\begin{tabular}{|c|c|c|c|}
				\hline
				$m_\mathrm{LQ}$\,[TeV]& Partons &$\sigma_\mathrm{S^{2/3}}$ [pb]&$\sigma_\mathrm{S^{4/3}}$ [pb]\\
				\hline
				\hline
				\multirow{9}{*}{0.9} & $d\ +\ {e}$ & $(8.85\times 10^{-2})^{+3.3\%}_{-3.7\%}\pm 2.0\%$ & $(9.21\times 10^{-2})^{+3.2\%}_{-3.5\%}\pm 2.0\%$ \\\cline{2-4}
				
				 &$d\ +\ {\mu}$ & $(8.54\times 10^{-2})^{+3.4\%}_{-3.7\%}\pm 2.0\%$ & $(8.90\times 10^{-2})^{+3.3\%}_{-3.6\%}\pm 2.0\%$ \\\cline{2-4}
				 
				 &$d\ +\ {\tau}$ & $(6.80\times 10^{-2})^{+4.0\%}_{-3.9\%}\pm 2.1\%$ & $(7.15\times 10^{-2})^{+3.9\%}_{-3.7\%}\pm 2.1\%$ \\
				 \hhline{|~|===|}
				 
				 &$s\ +\ {e}$& $(2.41\times 10^{-2})^{+3.8\%}_{-4.2\%}\pm 5.4\%$ & $(2.51\times 10^{-2})^{+3.7\%}_{-4.1\%}\pm 5.4\%$ \\\cline{2-4}
				 
				 &$s\ +\ {\mu}$ & $(2.34\times 10^{-2})^{+3.9\%}_{-4.3\%}\pm 5.4\%$ & $(2.44\times 10^{-2})^{+3.8\%}_{-4.1\%}\pm 5.4\%$ \\ \cline{2-4}
				 
				 &$s\ +\ {\tau}$ & $(1.85\times 10^{-2})^{+4.3\%}_{-4.5\%}\pm 5.5\%$ & $(1.95\times 10^{-2})^{+4.2\%}_{-4.3\%}\pm 5.5\%$ \\ \hhline{|~|===|}

				 &$b\ +\ {e}$ & $(9.01\times 10^{-3})^{+4.9\%}_{-5.8\%}\pm 1.8\%$ & $(9.39\times 10^{-3})^{+4.7\%}_{-5.6\%}\pm 1.8\%$ \\ \cline{2-4}
				 
				 & $b\ +\ {\mu}$ & $(8.76\times 10^{-3})^{+4.9\%}_{-5.8\%}\pm 1.8\%$ & $(9.14\times 10^{-3})^{+4.7\%}_{-5.6\%}\pm 1.8\%$ \\ \cline{2-4}
				 
				 &$b\ +\ {\tau}$ & $(6.87\times 10^{-3})^{+5.3\%}_{-6.3\%}\pm 2.1\%$ & $(7.25\times 10^{-3})^{+5.0\%}_{-5.9\%}\pm 2.0\%$ \\
				\hline
				\hline
				\multirow{9}{*}{1.6} & $d\ +\ {e}$ & $(7.23\times 10^{-3})^{+3.0\%}_{-3.2\%}\pm 2.3\%$ & $(7.47\times 10^{-3})^{+2.8\%}_{-3.1\%}\pm 2.3\%$ \\\cline{2-4}
				
				 &$d\ +\ {\mu}$ & $(7.07\times 10^{-3})^{+3.0\%}_{-3.3\%}\pm 2.3\%$ & $(7.30\times 10^{-3})^{+2.9\%}_{-3.1\%}\pm 2.3\%$ \\ \cline{2-4}
				 
				 &$d\ +\ {\tau}$ & $(5.76\times 10^{-3})^{+3.7\%}_{-3.4\%}\pm 2.4\%$ & $(6.00\times 10^{-3})^{+3.5\%}_{-3.3\%}\pm 2.4\%$ \\ \hhline{|~|===|}
				 
				 &$s\ +\ {e}$& $(1.40\times 10^{-3})^{+3.5\%}_{-3.7\%}\pm 8.8\%$ & $(1.44\times 10^{-3})^{+3.4\%}_{-3.5\%}\pm 8.8\%$ \\ \cline{2-4}
				 
				 &$s\ +\ {\mu}$ & $(1.37\times 10^{-3})^{+3.5\%}_{-3.7\%}\pm 8.7\%$ & $(1.42\times 10^{-3})^{+3.4\%}_{-3.6\%}\pm 8.7\%$ \\ \cline{2-4}
				 
				 &$s\ +\ {\tau}$ & $(1.12\times 10^{-3})^{+4.1\%}_{-3.9\%}\pm 8.8\%$ & $(1.16\times 10^{-3})^{+4.0\%}_{-3.7\%}\pm 8.8\%$ \\ \hhline{|~|===|}
				 
				 &$b\ +\ {e}$ & $(4.40\times 10^{-4})^{+4.5\%}_{-5.1\%}\pm 2.4\%$ & $(4.55\times 10^{-4})^{+4.3\%}_{-5.0\%}\pm 2.4\%$ \\ \cline{2-4}
				 
				 & $b\ +\ {\mu}$ & $(4.32\times 10^{-4})^{+4.5\%}_{-5.1\%}\pm 2.4\%$ & $(4.47\times 10^{-4})^{+4.4\%}_{-5.0\%}\pm 2.4\%$ \\ \cline{2-4}
				 
				 &$b\ +\ {\tau}$ & $(3.49\times 10^{-4})^{+4.8\%}_{-5.5\%}\pm 2.6\%$ & $(3.63\times 10^{-4})^{+4.6\%}_{-5.3\%}\pm 2.6\%$ \\
				\hline
				\hline
				\multirow{9}{*}{2.5} & $d\ +\ {e}$ & $(6.63\times 10^{-4})^{+2.6\%}_{-2.9\%}\pm 2.9\%$ & $(6.80\times 10^{-4})^{+2.4\%}_{-2.8\%}\pm 2.9\%$ \\\cline{2-4}
				
				 &$d\ +\ {\mu}$ & $(6.54\times 10^{-4})^{+2.6\%}_{-2.9\%}\pm 2.9\%$ & $(6.71\times 10^{-4})^{+2.5\%}_{-2.8\%}\pm 2.9\%$ \\ \cline{2-4}
				 
				 &$d\ +\ {\tau}$ & $(5.43\times 10^{-4})^{+3.3\%}_{-3.0\%}\pm 3.0\%$ & $(5.60\times 10^{-4})^{+3.1\%}_{-2.9\%}\pm 3.0\%$ \\ \hhline{|~|===|}
				 
				 &$s\ +\ {e}$& $(9.66\times 10^{-5})^{+2.9\%}_{-3.3\%}\pm 16.0\%$ & $(9.90\times 10^{-5})^{+2.8\%}_{-3.2\%}\pm 16.0\%$ \\ \cline{2-4}
				 
				 &$s\ +\ {\mu}$ & $(9.53\times 10^{-5})^{+3.0\%}_{-3.2\%}\pm 15.9\%$ & $(9.77\times 10^{-5})^{+2.9\%}_{-3.2\%}\pm 15.9\%$ \\ \cline{2-4}
				 
				 &$s\ +\ {\tau}$ & $(7.89\times 10^{-5})^{+3.6\%}_{-3.4\%}\pm 16.0\%$ & $(8.14\times 10^{-5})^{+3.4\%}_{-3.3\%}\pm 16.0\%$ \\ \hhline{|~|===|}
				 
				 &$b\ +\ {e}$ & $(2.42\times 10^{-5})^{+4.2\%}_{-4.7\%}\pm 3.8\%$ & $(2.48\times 10^{-5})^{+4.2\%}_{-4.5\%}\pm 3.8\%$ \\ \cline{2-4}
				 
				 & $b\ +\ {\mu}$ & $(2.39\times 10^{-5})^{+4.3\%}_{-4.7\%}\pm 3.8\%$ & $(2.45\times 10^{-5})^{+4.2\%}_{-4.6\%}\pm 3.8\%$ \\ \cline{2-4}
				 
				 &$b\ +\ {\tau}$ & $(1.97\times 10^{-5})^{+4.6\%}_{-5.0\%}\pm 3.9\%$ & $(2.03\times 10^{-5})^{+4.5\%}_{-4.8\%}\pm 3.9\%$ \\
				\hline
				\hline
				\multirow{9}{*}{4.0} & $d\ +\ {e}$ & $(2.41\times 10^{-5})^{+2.1\%}_{-2.5\%}\pm 4.7\%$ & $(2.45\times 10^{-5})^{+2.1\%}_{-2.4\%}\pm 4.7\%$ \\\cline{2-4}
				
				 &$d\ +\ {\mu}$ & $(2.39\times 10^{-5})^{+2.1\%}_{-2.5\%}\pm 4.7\%$ & $(2.43\times 10^{-5})^{+2.1\%}_{-2.4\%}\pm 4.7\%$ \\\cline{2-4}
				 
				 &$d\ +\ {\tau}$ & $(2.02\times 10^{-5})^{+2.6\%}_{-2.6\%}\pm 4.7\%$ & $(2.06\times 10^{-5})^{+2.4\%}_{-2.5\%}\pm 4.7\%$ \\\hhline{|~|===|}
				 
				 &$s\ +\ {e}$& $(2.84\times 10^{-6})^{+2.4\%}_{-2.8\%}\pm 37.5\%$ & $(2.89\times 10^{-6})^{+2.4\%}_{-2.8\%}\pm 37.7\%$ \\\cline{2-4}
				 
				 &$s\ +\ {\mu}$ & $(2.81\times 10^{-6})^{+2.4\%}_{-2.8\%}\pm 37.5\%$ & $(2.87\times 10^{-6})^{+2.4\%}_{-2.8\%}\pm 37.6\%$ \\\cline{2-4}
				 
				 &$s\ +\ {\tau}$ & $(2.37\times 10^{-6})^{+2.5\%}_{-3.0\%}\pm 37.5\%$ & $(2.42\times 10^{-6})^{+2.5\%}_{-2.9\%}\pm 37.7\%$ \\\hhline{|~|===|}
				 
				 &$b\ +\ {e}$ & $(4.32\times 10^{-7})^{+3.9\%}_{-4.1\%}\pm 10.2\%$ & $(4.40\times 10^{-7})^{+3.8\%}_{-4.1\%}\pm 10.2\%$ \\\cline{2-4}
				 
				 & $b\ +\ {\mu}$ & $(4.29\times 10^{-7})^{+3.9\%}_{-4.2\%}\pm 10.2\%$ & $(4.37\times 10^{-7})^{+3.8\%}_{-4.1\%}\pm 10.2\%$ \\\cline{2-4}
				 
				 &$b\ +\ {\tau}$ & $(3.59\times 10^{-7})^{+4.5\%}_{-4.4\%}\pm 10.3\%$ & $(3.67\times 10^{-7})^{+4.3\%}_{-4.3\%}\pm 10.3\%$ \\
				\hline
\end{tabular}}
\caption{\label{tab:13down} Inclusive cross sections in {\bf pb} for the resonant leptoquark production from down-type quarks, $p p \to $\,LQ + charge-conjugated process, as a function of the leptoquark mass $m_{{\rm LQ}}$ at $\sqrt{s}=13$\,TeV. The cross section $\sigma_\mathrm{S^{2/3}}\; (\sigma_\mathrm{S^{2/3}})$ corresponds to the resonant production of scalar LQ with absolute electric charge $2/3\ (4/3)$ when the associated Yukawa coupling strength is set to one, $y_{q\ell} = 1$. The second column denotes which quark-lepton pair couples to the corresponding leptoquark. First (second) uncertainty is due to the renormalisation and factorisation scale variations (PDF replicas),  and is given in per cent units. See Section~\ref{sec:dis} for details.}
\end{table}

Partonic cross sections are convoluted with {\tt LUXlep-NNPDF31\_nlo\_as\_0118\_luxqed} (v2) parton distribution functions derived in Ref.~\cite{Buonocore:2020nai}. To this purpose, we employ the Mathematica package {\tt ManeParse}~\cite{Clark:2016jgm} for manipulating the {\tt LHAPDF} grids~\cite{Buckley:2014ana}. The PDF extrapolation in $Q^2$ is checked by solving the corresponding DGLAP equations using {\tt Hoppet}~\cite{Salam:2008qg} in accordance with the prescription from~\cite{Buonocore:2020nai}. Also, the running of the gauge couplings with the renormalisation scale is appropriately included. The central renormalisation and factorisation scales are set to $\mu_{\rm R}=\mu_{\rm F}=m_{{\rm LQ}}$. We estimate the uncertainty from the missing higher order corrections by varying the scales in the range $\{\mu_{\rm R},\mu_{\rm F}\} \in [0.5 - 2]\,m_{{\rm LQ}}$\,, while respecting $0.5 \leq \mu_{\rm R}/\mu_{\rm F} \leq 2$. Independently of $\mu_{\rm R}$ and $\mu_{\rm F}$ scale variations, the renormalisation group running of the leptoquark coupling $y_{q\ell}(\mu)$ in the range $\mu \in [0.5 - 2]~m_{{\rm LQ}}$ leads to the cross section prediction uncertainty of about $4\%$ across the entire $m_{{\rm LQ}}$ window. The uncertainties due to the parton distribution functions are calculated by the method of replicas~\cite{Ball:2017nwa,Bertone:2017bme}. In particular, we report the standard deviation of the result calculated over one hundred replicas as the PDF error.

The lepton and antilepton PDFs are numerically the same. However, this is not the case for the light quarks, implying that e.g. $u e^-$ induced cross section is different from $\bar u e^+$. We therefore report the cross sections for $p p \to$\,LQ + the charge-conjugated process in Table~\ref{tab:13up} (up-type quarks) and Table~\ref{tab:13down} (down-type quarks), at NLO QCD + QED accuracy. Thanks to the inclusion of radiative corrections computed in Section~\ref{sec:scalarLQ}, the uncertainties due to the $\{\mu_{\rm R},\mu_{\rm F}\}$ scale variations are at the level of few per cent for all leptoquark charges, as well as, quark and lepton flavors and benchmark masses. The uncertainties due to the parton distribution functions strongly depend on the quark flavor and the leptoquark mass. In particular, the total uncertainty becomes dominated by the limited knowledge of the heavy quark PDFs when $m_{{\rm LQ}}$ is several TeV.

\begin{figure}[tbp]
\centering
\includegraphics[scale=1.2]{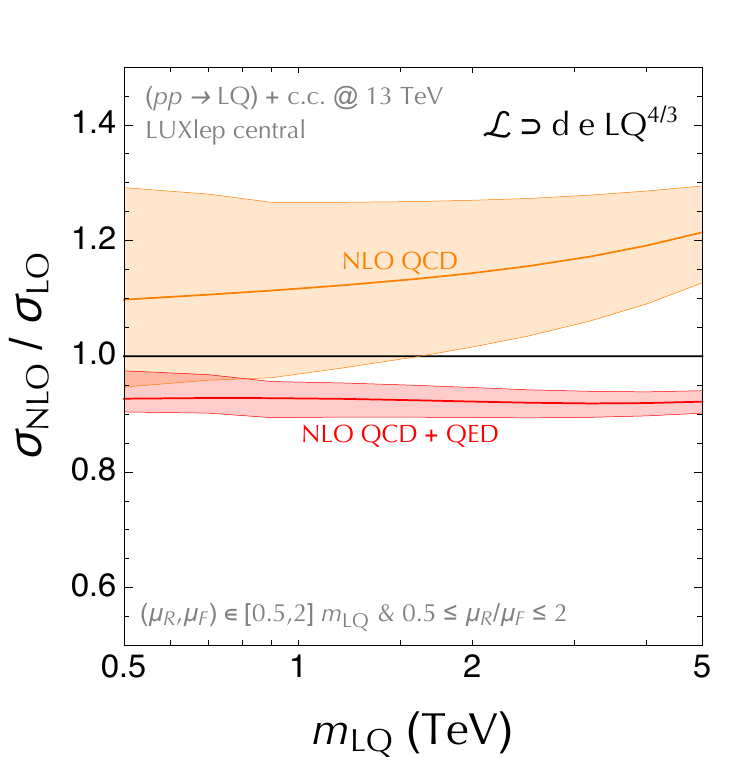} 
\caption{\label{fig:SVelectronMain} NLO $K$-factor ($\sigma_{{\rm NLO}} / \sigma_{{\rm LO}}$) for the resonant scalar leptoquark production at 13~TeV LHC. Shown with orange (red) solid lines are the NLO QCD (NLO QCD + QED) predictions normalised to the LO when setting the central scales to $\mu_{\rm R}=\mu_{\rm F}=m_\mathrm{LQ}$. The colored bands are obtained by varying factorisation and renormalisation scales in the NLO calculations within $\{ \mu_{\rm F}, \mu_{\rm R} \} \in [0.5, 2]~ m_\mathrm{LQ}$ while respecting $1/2 \le \mu_{\rm R}/\mu_{\rm F} \le 2$. We use the central PDF set from {\tt LUXlep-NNPDF31\_nlo\_as\_0118\_luxqed} (v2)~\cite{Buonocore:2020nai}. The benchmark example in the plot is a leptoquark with $|Q_{{\rm LQ}}| = 4/3$ coupled to the down quark and electron. All other cases are shown in Figs.~\ref{fig:SVelectron},~\ref{fig:SVmuon}, and~\ref{fig:SVtau} (see Appendix~\ref{app:supplement}).}
\end{figure}

Next-to-leading order $K$-factors, defined as the ratio of NLO to LO results, are shown in Appendix~\ref{app:supplement} in Figs.~\ref{fig:SVelectron}\,(electron),~\ref{fig:SVmuon}\,(muon), and~\ref{fig:SVtau}\,(tau) for all possible quark and lepton flavors and leptoquark charges. One notable example is shown in Fig.~\ref{fig:SVelectronMain} in the main text.  These are calculated using the central PDF set and the central scales $\mu_{\rm R}=\mu_{\rm F}=m_{{\rm LQ}}$ for the LO cross section, while at NLO, we consider $\{ \mu_{\rm R},\mu_{\rm F} \}$ scale variation with the central PDF set. The red (orange) bands are with (without) NLO QED corrections.\footnote{Our calculation is also an important test of the lepton PDFs derived in Ref.~\cite{Buonocore:2020nai}.} In all cases considered, the error band dramatically shrinks, illustrating the importance of the NLO QED corrections. Interestingly, both QCD and QED corrections are large, however, they partially cancel in the total cross section. Inspecting Figs.~\ref{fig:SVelectron},~\ref{fig:SVmuon}, and~\ref{fig:SVtau} we conclude that $K$-factors typically exhibit only a slight dependence on the leptoquark mass and electric charge, as well as, lepton flavors. In this calculation, we sum up cross sections for the process $p p \to $\,LQ and the charge-conjugated process before taking the ratio. We checked that the individual $K$-factors for the two are very close to each other, thus we report only the $K$-factors for the sum. 

We also study the dependence of the NLO $K$-factors on the PDF uncertainties. In particular, for every PDF replica we compute $\sigma_{{\rm NLO}}/\sigma_{{\rm LO}}$. We then derive the $68\%$ confidence level range around the central PDF prediction. Interestingly, this band does not exceed the NLO QCD~+~QED scale variation band, except for a very heavy leptoquark close to the edge of the considered mass range, where the PDF errors are $\mathcal{O}(1)$ for some flavors. In other words, PDF uncertainties cancel in the ratio to a good approximation. We therefore conclude that the $K$-factors reported in Appendix~\ref{app:supplement} are robust, and will not change significantly by more precise PDFs in the future.

On the practical side, the existing leading order generators are missing the leptonic shower crucial to properly simulate the resonant leptoquark events. However, this shortcoming will soon be resolved, see the third footnote in Ref.~\cite{Buonocore:2020erb}. Once this is in place, the $K$-factors derived in this paper can be directly applied to the future LHC resonant leptoquark searches to correct the overall signal yield. The main experimental difference between the single leptoquark plus lepton production and the resonant leptoquark production is the $p_T$ spectrum of the accompanied lepton. In particular, the lepton is hard (soft) in the former (latter) case. Therefore, measuring the lepton (or the leptoquark) $p_T$ distribution will enable efficient discrimination between different leptoquark production mechanisms at the LHC. To this purpose, it is crucial to have a good theoretical control over the $p_T$ spectrum. Our Appendix could serve as a starting point for this calculation. Note, that the leptoquark searches so far required the presence of two charged leptons which effectively vetos the resonant mechanism.

The leptoquark signature is quite unique; it will show up as a resonance in the jet-lepton invariant mass distribution.  To study the flavor structure of the underlying interactions, one can make use of the flavor tagging of the decay products. Unfortunately, on the quark side there is a big degeneracy among light quarks $u,d,s$ and $c$ which are somewhat distinguished from the $b$ quark. The task is even more difficult on the production side. The ratio of the rates for the single leptoquark plus lepton production and the resonant leptoquark production does not depend on the value of the leptoquark Yukawa coupling, however, it is sensitive to the initial quark flavor. This can be used to determine the flavor structure of the dominant leptoquark coupling in production. We have checked that the ratio drops quickly with the leptoquark mass and the discrepancy is more pronounced for sea quarks than for valence quarks. 

Another observable relevant for the leptoquark flavor physics in high-$p_T$ collisions is the ratio of the resonant rate $p p \to$\,LQ to its charge-conjugated process. For heavy $c$ and $b$ quarks the two rates are the same, while for the valence quarks the two rates can differ by a factor of $\mathcal{O}(10)$. We have checked that this observable indeed has a discriminating power, however, a dedicated analysis is needed to make a quantitative statement. We also noticed a large PDF uncertainties in the prediction of this ratio attributed to the poor knowledge of sea quarks at large $x$. Therefore, the success of this method depends on the improvements in measuring sea quark parton distribution functions.

\newpage	
%%%%%%%%%%%%%%%%%%%%%%%%%%%%%%%%%%%%%%
\section{Conclusions}
\label{sec:Conc}
%%%%%%%%%%%%%%%%%%%%%%%%%%%%%%%%%%%%%%

A discovery of a leptoquark at the Large Hadron Collider would fundamentally change our understanding of particle physics, pointing towards a microscopic theory where quarks and leptons unify. Viable extensions of the Standard Model with TeV-scale leptoquarks exist, and are safe on proton decay and dangerous flavor changing neutral currents. Moreover, these models have recently received a large attention within the community. Namely, leptoquarks in the TeV mass range provide an elegant explanation of the {long-standing} hints on the lepton flavor universality violation in $B$-meson decays, as well as, the anomalous magnetic moment of the muon. 

Leptoquark collider searches so far were mainly focused on the pair production mechanism driven by QCD interactions, while the role of the defining leptoquark interaction to a quark and a lepton was invoked in decays. However, interesting flavor effects occur when the leptoquark coupling is large(ish)~\cite{Buttazzo:2017ixm,Angelescu:2018tyl,Aebischer:2019mlg,Dorsner:2019itg,Gherardi:2020qhc,Aaboud:2017efa,Mandal:2019gff,Bordone:2020lnb,Fuentes-Martin:2019mun,Crivellin:2020mjs,Altmannshofer:2020ywf,Bause:2020xzj,Brdar:2020quo,Bause:2019vpr}, consequently predicting richer collider phenomenology on the production side. Building on Refs.~\cite{Buonocore:2020erb,Buonocore:2020nai}, in this paper we study {\it the resonant leptoquark} production mechanism. Namely, the quantum fluctuations allow for a small presence of a lepton inside the proton which fuses with a quark from the other proton, to produce a leptoquark. The smallness of the lepton distribution is overcome by the resonant enhancement, providing this mechanism with the largest cross sections of all when $m_{{\rm LQ}} \gtrsim 1$~TeV and $y_{q\ell}\sim\mathcal{O}(1)$, see Fig~\ref{fig:xsecComparison}.

We calculate for the first time next-to-leading order QCD and QED corrections to the resonant leptoquark production at hadron colliders. The present study is limited to scalar leptoquarks while the vector leptoquark case is left for the future work. The total cross section is given in closed form in Eqs.~\eqref{eq:xsecsve},~\eqref{eq:xsecql},~\eqref{eq:xsecgl}, and~\eqref{eq:xsecqgamma}, and the detailed derivation is carried out in Appendices~\ref{sec:QEDNLO} and~\ref{sec:NLOQCD}. This formula is numerically integrated with the most recent lepton PDFs~\cite{Buonocore:2020nai} to obtain the hadronic cross sections at the LHC. The main numerical results are reported in Tables~\ref{tab:13up} and~\ref{tab:13down}, and in Figs.~\ref{fig:SVelectron},~\ref{fig:SVmuon} and~\ref{fig:SVtau}. The calculation is performed for a set of benchmark points in the mass range relevant for the future searches, as well as, for all possible lepton and quark flavors and leptoquark charges. Importantly, our results are applicable for a general scalar leptoquark model with arbitrary flavor couplings. 

We find that both QCD and QED corrections are large and are of similar size. However, they come with the opposite sign and cancel out in the final cross section, leading to somewhat smaller corrections of the tree-level result than initially expected. However, the advantage of our calculation is that we are now in position to reliably estimate the theoretical uncertainties. On this note, we observed a dramatic reduction of the renormalisation and factorisation scale variation uncertainties after the inclusion of QED corrections on top of the QCD ones. This is nicely illustrated in Fig.~\ref{fig:SVelectronMain} with the red band. The leading source of theoretical error at this point is the limited knowledge of the parton distribution functions, in particular, the sea quark PDFs at large $x$. The breakdown of different uncertainties is summarised in the predictions for the total cross sections in Tables~\ref{tab:13up} and~\ref{tab:13down}. The complete set of NLO $K$-factors is reported in Appendix~\ref{app:supplement} and can be straightforwardly applied in the future experimental searches at the LHC for the most general leptoquark model.

Finally, should a leptoquark be discovered at the LHC, precision measurements of the resonant process and its charge-conjugate, as well as, the single leptoquark plus lepton production, would help to deduce the flavor of the leptoquark interactions. Hopefully, synchronised deviations would show up in the low-energy flavor transitions to confirm this picture.

\acknowledgments

We thank Gino Isidori, Javier Fuentes-Mart\'{\i}n, Matthias K\"onig and Pier Francesco Monni for useful discussions. The work of AG has received funding from the Swiss National Science Foundation (SNF) through the Eccellenza Professorial Fellowship ``Flavor Physics at the High Energy Frontier'' project number 186866. The work of AG and NS is supported by the European Research Council (ERC) under the European Union’s Horizon 2020 research and innovation programme, grant agreement 833280 (FLAY).

\appendix
	
%%%%%%%%%%%%%%%%%%%%%%%%%%%%%%%%%%%%%%
\section{NLO QED corrections to resonant production}
\label{sec:QEDNLO}
%%%%%%%%%%%%%%%%%%%%%%%%%%%%%%%%%%%%%%
	
\begin{figure}[H]
\centering
\includegraphics[width=1\textwidth]{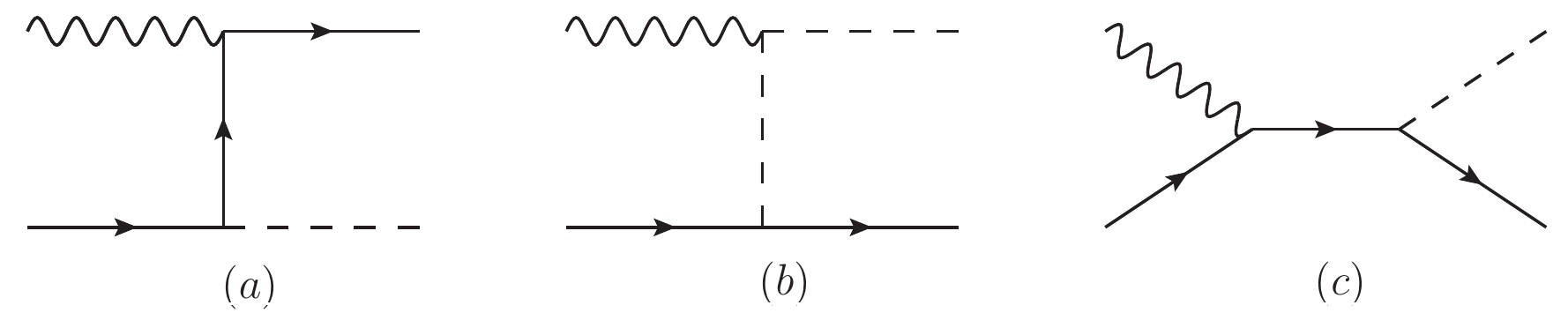}
\vspace{-0.6cm}
\caption{Diagrams for the process $\gamma + q\to l^- + LQ$ contributing to the resonant leptoquark production at $\mathcal{O}(\alpha)$.}
\label{fig:QEDdown}
\end{figure}

The QED corrections to the resonant leptoquark production correspond to processes involving a photon in the initial state splitting into a lepton pair. As explained in Section~\ref{sec:QEDmain} , the inclusion of these corrections is necessary for calculating the resonant production at $\mathcal{O}(\alpha_s^3)$, which is a typical size of the NLO QCD corrections. Due to different electric charges, the QED corrections to production cross section will differ for various leptoquark types. We can resonantly produce all types of scalar leptoquarks using the suitable partons inside the colliding protons. The possible combinations are
\begin{equation}
\begin{aligned}
\bar{u}+e^+ &\longrightarrow S_{1/3}\,,\\
d+e^+ &\longrightarrow S_{2/3}\,,\\
\bar{d}+e^+ &\longrightarrow S_{4/3}\,,\\
u+e^+ &\longrightarrow S_{5/3}\,,
\end{aligned}
\label{eq:comb}
\end{equation}
along with the corresponding charge conjugated processes, where the fraction in the subscript denotes the leptoquark electric charge. The QED correction for each of the listed process is given by three diagrams shown in Fig. \ref{fig:QEDdown}. The initial state quark and photon create the complementary scalar leptoquark together with a soft charged lepton in the final state. 
	
\noindent Generically, the amplitude for the process  $\gamma(p_1) + q(p_2) \to l^-(k) + LQ^+(q)$ obtained by interfering diagrams reads
\begin{equation}
i\mathcal{M} = - i y_{q\ell}\ e\  \bar{u}(k) \text{P}_{L,R} \left[\gamma^\mu\frac{\left(\slashed{p_1}-\slashed{k}\right)}{\left(p_1-k\right)^2} + Q_{\rm LQ}\frac{\left(2q-p_1\right)^\mu}{\left(q-p_1\right)^2-m^2} + Q_{\rm q} \frac{\left(\slashed{p_1}+\slashed{p_2}\right)}{\left(p_1+p_2\right)^2}\gamma^\mu\right]\zeta(p_2)\epsilon_\mu(p_1)\,,
\end{equation}
where $p_1\ (p_2)$ is the four-momentum of the photon ((anti-)quark) in the initial state, while $q\ (k)$ is the four-momentum of the leptoquark (soft charged lepton) in the final state. The fermionic wave-function $\zeta(p_2)$ could either stand for the particle or anti-particle, depending on the produced leptoquark type. The partonic cross section calculation presented is the same for all types of leptoquarks, with the only difference provided by different particle charges, $Q_{\rm q}$ denoting the (anti-)quark and $Q_{\rm LQ}$ the leptoquark charge.   To express the kinematics, it is convenient to use the center of mass frame in which
\begin{equation}
p_1^\mu = \frac{\sqrt{\hat{s}}}{2}(1,0,0,1),\quad p_2^\mu = \frac{\sqrt{\hat{s}}}{2}(1,0,0,-1),\quad
k = \frac{\sqrt{\hat{s}}}{2}(1-z)(1,0,\sin{\theta},\cos{\theta})\,,
\end{equation}
with $z=m^2_{{\rm LQ}}/\hat{s}$. Additionally, the relation between partonic Mandelstam variables $\hat{s}=(p_1+p_2)^2$ and\ \  $\hat{t}=(p_1-k)^2$\ \  becomes\ \ $\hat{t}= - \hat{s}w(1-z)$, where $w=(1-\cos{\theta})/2$. The collinear divergences that appear when soft lepton is emitted parallel to the photon are regulated using dimensional regularisation with $d=4-2\epsilon$. Averaging over the initial, and summing over the final polarisations and colors, the averaged squared matrix element can be written as
\begin{equation}
|\overline{\mathcal{M}}|^2 = \frac{|y_{q\ell}|^2 e^2}{d-2}\left(\mathcal{M}^2_{\text{div}} + \mathcal{M}^2_{\text{fin}}\right)\,,
\end{equation}
where $d-2$ in the denominator counts the polarisations of the massless gauge bosons in $d$-dimensions, and $\mathcal{M}^2_{\text{div}}$ ($\mathcal{M}^2_{\text{fin}}$) denotes the part of the averaged squared matrix element that will produce the IR-divergent (IR-finite) contributions to the cross section after integration over the phase-space. In terms of $w$ and $z$, they can be written as
\begin{alignat}{2}
\mathcal{M}^2_{\text{div}} &= &&\  \frac{1}{w}\left[\frac{d-2}{2(1-z)} +2 Q_{\rm q} z + Q_{\rm LQ}\left(1-\frac{1+2z}{1-w(1-z)}\right)\right],\\
\mathcal{M}^2_{\text{fin}} &= &&\ 
\frac{(1-w)(1-z)}{1-w(1-z)}\left[Q_{\rm LQ}^2\left(1-\frac{2z}{1-w(1-z)}\right)-Q_{\rm q}Q_{\rm LQ}\left(1-2z\right)\right]
\\\nonumber
&\ &&\ + Q_{\rm q}\left(d-2(1+z)\right) - Q_{\rm LQ}\left(1-\frac{1+2z}{1-w(1-z)} \right) + \frac{d-2}{2}Q_{\rm q}^2w(1-z).
\end{alignat}
Moreover, the integration over the 2-body phase-space in $d$-dimensions with the corresponding flux factor, expressed in terms of $w$ and $z$, can be performed as
\begin{equation} \frac{1}{16\pi\hat{s}}\left(\frac{4\pi\mu^2}{\hat{s}}\right)^{\frac{4-d}{2}}\int_0^1 |\overline{\mathcal{M}}|^2 \frac{\left[w(1-w)\right]^{\frac{d-4}{2}}(1-z)^{d-3}}{\Gamma(\frac{d-2}{2})}\ dw.
\end{equation}
The integral over $\mathcal{M}^2_{\text{fin}}$ is IR safe for $d=4$. The result for the finite contribution to the partonic cross section is
\begin{alignat}{2}
\label{eq:xsecfin}
\hat{\sigma}_{\text{fin}} &=\ \frac{\pi |y_{q\ell}|^2}{4\hat{s}}\frac{\alpha}{2\pi}\bigg[&&Q_{\rm q} Q_{\rm LQ}(1-2z)(z-z\log z-1) + Q_{\rm LQ}^2(1+z-2z^2+3z\log z)\\\nonumber
&\ &&\ + 2 Q_{\rm q} \left(1+\frac{Q_{\rm q}}{4}\right) (1-z)^2+Q_{\rm LQ} (z-(1+2z)\log z-1)\bigg].
\label{eq:finite}
\end{alignat}
On the other hand, the phase-space integration over $\mathcal{M}^2_{\text{div}}$ induces IR-poles in the partonic cross section. Regulating the integral we obtain
\begin{gather}
\hat{\sigma}_{\text{div}}  = \frac{\pi |y_{q\ell}|^2}{4\hat{s}}\frac{\alpha}{2\pi} \left(\frac{4\pi\mu^2}{\hat{s}}\right)^\epsilon\frac{1}{\Gamma(1-\epsilon)}\int w^{-1-\epsilon}(1-w)^{-\epsilon}(1-z)^{1-2\epsilon}\mathcal{F}(w,z)dw\,,\\
\mathcal{F}(w,z)  = \frac{1}{1-z} +(1+\epsilon)\left[2Q_{\rm q} z+ Q_{\rm LQ}\left(1-\frac{1+2z}{1-w(1-z)}\right)\right].
\end{gather}
The IR-pole in $\hat{\sigma}_{\text{div}}$ becomes explicit after $w^{-1-\epsilon}$ is expanded around $\epsilon=0$ to give a distribution
\begin{equation}
w^{-1-\epsilon} = -\frac{1}{\epsilon}\delta(w) + \frac{1}{w_+}+\mathcal{O}(\epsilon),
\end{equation}
with the plus distribution defined such that
\begin{equation}
\int_{0}^{1}\frac{f(1-w)}{w_+}\ dw = \int_{0}^{1}\frac{f(1-w)-f(1)}{w}\ dw.
\end{equation}
The contribution to the partonic cross section containing a collinear divergence is then
\begin{align}
\nonumber\hat{\sigma}_{\text{div}} & = \frac{\pi |y_{q\ell}|^2}{4\hat{s}}\frac{\alpha}{2\pi} \left(\frac{4\pi\mu^2}{\hat{s}}\right)^\epsilon\frac{1}{\Gamma(1-\epsilon)}\bigg[-\frac{1}{\epsilon}\bigg(1+2(Q_{\rm q}-Q_{\rm LQ})z(1-z)\bigg)+2\log (1-z)\\
&\quad + Q_{\rm LQ}(1-z)(1+2z)\log z - 2(Q_{\rm q} - Q_{\rm LQ})z(1-z)(1-2\log (1-z))\bigg].
\label{eq:div}
\end{align}
Note that the divergence is universal for all  combinations listed in \eqref{eq:comb} as $(Q_{\rm q}-Q_{\rm LQ}) = -1$, for all of them, and the corresponding coefficient is identified with the leading-order photon-to-charged lepton splitting function
\begin{equation}
P_{\ell \leftarrow \gamma}(z) = z^2+(1-z)^2.
\end{equation}
	
\noindent In order to calculate the measured hadronic cross section for these processes, we need to convolute the partonic cross section with the corresponding quark and photon PDFs. In the procedure, the  collinear singularity can be absorbed into the bare PDFs at a factorisation scale $\mu_{\rm F}$.  Here, we utilize the $\overline{\rm MS}$ factorisation scheme by adding the counter term 
\begin{equation}
\hat{\sigma}^{CT} =\frac{\pi |y_{q\ell}|^2}{4\hat{s}}\frac{\alpha}{2\pi}(4\pi)^\epsilon \frac{1}{\epsilon\Gamma(1-\epsilon)} P_{\ell \leftarrow \gamma}(z),
\label{eq:ct}
\end{equation}
to the partonic cross section, consistent with the $\overline{\rm MS}$ prescription used in extracting parton density functions presented in~\cite{Buonocore:2020nai}.  Combining relations \eqref{eq:finite}, \eqref{eq:div} and \eqref{eq:ct}, we find that the result is finite, factorisation scale dependent, and can be written in the following form
\begin{equation}
\hat{\sigma}_{q \gamma}(z) = \frac{\pi z |y_{q\ell}|^2}{4m_{\rm LQ}^2}\frac{\alpha}{2\pi} \left(-\log\left(\frac{\mu_{\rm F}^2}{\hat{s}(1-z)^2}\right)P_{\ell \leftarrow \gamma}(z)+X_{Q_{LQ}}(z)\right).
\label{eq:gammad}
\end{equation}
The non-universal pieces for different scalar leptoquarks, after the electric charges are replaced, read 
\begin{align}
X_{1/3}(z) &=  - \frac{2}{9}(1-z)(5-13z)
+ \frac{2}{9}(1-5z)z\log z,\\
X_{2/3}(z) &= - \frac{11}{18}(1-z)(1-5z)
+ \frac{8}{9}(1-2z)z\log z,\\
X_{4/3}(z) &=  \frac{1}{18}(1-z)(13+103z)
+ \frac{16}{9}(2-z)z\log z,\\
X_{5/3}(z) &=   \frac{2}{9}(1-z)(7+37z)
+ \frac{10}{9}(5-z)z\log z,
\end{align}
where $\{1/3,\ 2/3,\ 4/3,\ 5/3\}$ correspond to electric charge of the leptoquark in the final state.

\section{NLO QCD corrections to resonant production}
\label{sec:NLOQCD}
	
\subsection{Virtual QCD corrections}
	
\begin{figure}[H]
\centering
\includegraphics[width=0.69\textwidth]{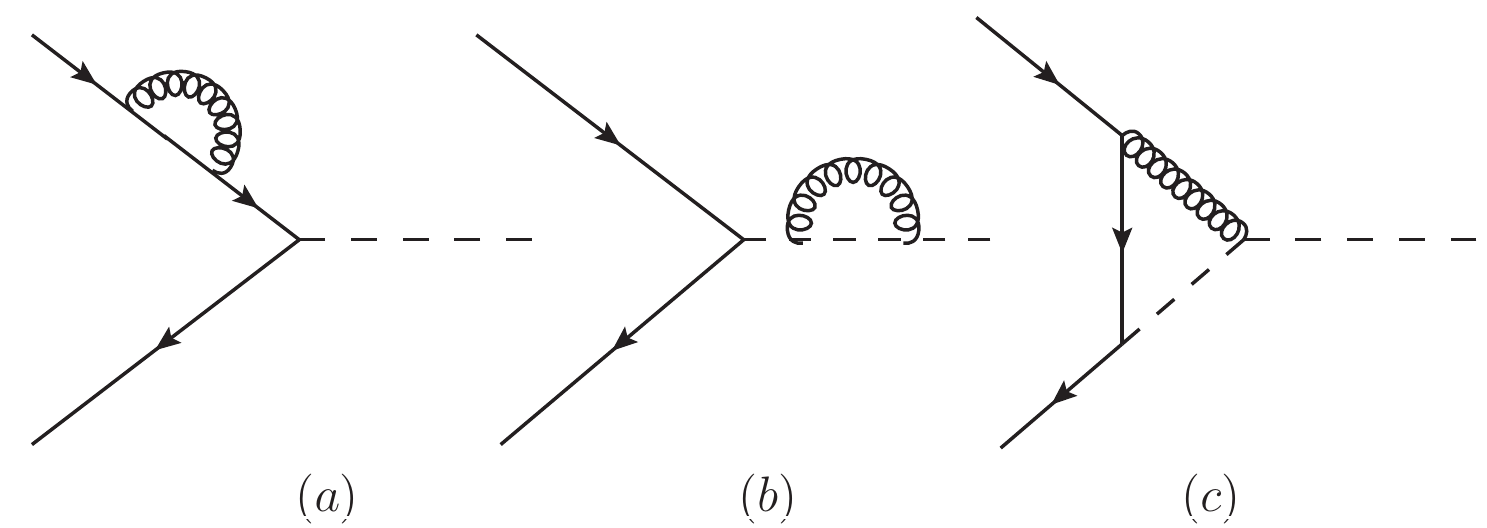}
\caption{The virtual corrections to the resonant leptoquark production at $\mathcal{O}(\alpha_s)$.}
\label{fig:QCDvirtual}
\end{figure}

The calculation of the $\mathcal{O}(\alpha_s)$ virtual corrections to the resonant leptoquark production proceeds as shown in Fig.~\ref{fig:QCDvirtual}: (a) the quark wave-function correction, (b) the leptoquark wave-function correction, and (c) the vertex correction. The UV finiteness of the results is achieved by adding the $\overline{\text{MS}}$ counter-terms.
	
\noindent Adding the virtual contributions to the tree-level amplitude, the NLO amplitude may be written as
\begin{equation}
\mathcal{A}_{\text{NLO}}=\mathcal{A}_{\text{tree}}\left[1+\frac{\alpha_s}{4\pi}\left(\frac{1}{2}\delta Z_{\rm q}(0)+\frac{1}{2}\delta Z_{\rm LQ}(m_{\rm LQ}^2)+\delta V_{\rm LQ}(m_{\rm LQ}^2)\right)\right]\ .
\label{eq:ANLO}
\end{equation}
The chiral Yukawa couplings result in the vanishing leptoquark contribution to the massless fermion wave-function, with gluon providing the only contribution
\begin{equation}
\delta Z_{\rm q}(0) = C_F\left(\frac{1}{\epsilon_{\rm IR}}+L_{\mu}^{\rm IR}-L_{\mu}^{\rm UV}\right)\,,
\end{equation}
where $L_{\mu}^{\rm IR(UV)} = \log(\mu_{\rm F(R)}^2/m_{\rm LQ}^2)$, with $\mu_{\rm F}$ and $\mu_{\rm R}$ denoting the factorisation and renormalisation scales, respectively, and $C_F=4/3$.  Similarly, the leptoquark two-point function $\Sigma_{LQ}(q^2)$, with $q$ being the leptoquark four-momentum, receives no contribution from fermions, and the only effect is caused by the leptoquark coupling to gluons. We renormalise the leptoquark mass on-shell, with the wave-function correction defined as
\begin{gather}
\delta Z_{\rm LQ}(q^2) = \frac{\Sigma_{\rm LQ}(q^2)-\Sigma_{\rm LQ}(m_{\rm LQ}^2)}{q^2-m_{\rm LQ}^2}\ .
\end{gather}
Taking the on-shell limit for the resonant leptoquark production $q^2= m_{\rm LQ}^2$, the correction becomes
\begin{equation}
\lim_{q^2\to m_{\rm LQ}^2}\delta Z_{\rm LQ}(q^2) = 2C_F\left(L_{\mu}^{\rm UV}-\frac{1}{\epsilon_{\rm IR}}-L_{\mu}^{\rm IR}\right)\,,
\end{equation}
while the vertex correction, evaluated on-shell reads
\begin{equation}
\delta V_{\rm LQ}(m_{\rm LQ}^2) = C_F\left[L_{\mu}^{\rm UV}-2\left(\frac{1}{\epsilon_{\rm IR}}+L_{\mu}^{\rm IR}\right)-\frac{1}{\epsilon_{\rm IR}^2}-\frac{1}{\epsilon_{\rm IR}}L_{\mu}^{\rm IR}-\frac{1}{2}\left(L_{\mu}^{\rm IR}\right)^2-\frac{\pi^2}{12}-2\right]\ .
\end{equation}
Combining the individual contributions listed above and integrating the averaged matrix element $|\overline{\mathcal{A}_{\text{NLO}}}|^2$ over the leptoquark phase-space, we obtain the virtual correction to the partonic cross section for the leptoquark resonant production
\begin{align}
\label{eq:virtual}
\hat{\sigma}^V(z) = \frac{\pi z |y_{q\ell}|^2}{4m_{\rm LQ}^2}\Bigg\{1+\frac{\alpha_s}{2\pi}C_F&\Bigg[\frac{3}{2}L_\mu^{\rm UV} - \frac{5}{2}\left(\frac{1}{\epsilon_{\rm IR}}+L_\mu^{\rm IR}\right)-\frac{1}{\epsilon_{\rm IR}^2}-\frac{1}{\epsilon_{\rm IR}}L_{\mu}^{\rm IR}\\\nonumber
&-\frac{1}{2}\left(L_{\mu}^{\rm IR}\right)^2-\frac{\pi^2}{12}-2\Bigg]\Bigg\}\ \delta(1-z).
\end{align}
    
\subsection{Real QCD corrections}
\begin{figure}[H]
\centering
\includegraphics[width=0.69\textwidth]{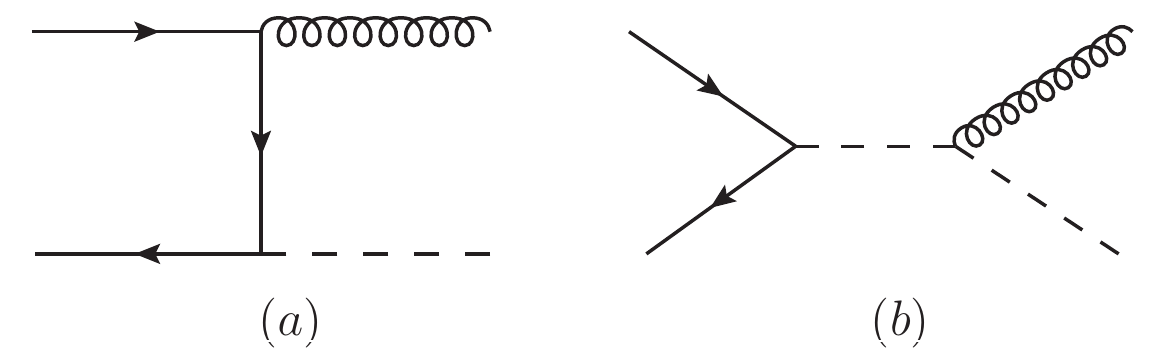}
\caption{Diagrams for the process $q + l\to g + LQ$ contributing to the resonant leptoquark production at $\mathcal{O}(\alpha_s)$.}
\label{fig:QCDquark}
\end{figure}

The calculation of the real QCD corrections closely follows the steps described in Appendix~\ref{sec:QEDNLO} for the QED case. We note that the results of this calculation already exist in the literature~\cite{Kunszt:1997at,Plehn:1997az} (see also~\cite{Djouadi:1989md}), which we have checked and found complete agreement. With that in mind, here we present the corresponding results, and commit to these references for more details. The first process to consider is the one with the soft gluon in the final state, which can happen either by emission from the quark or the leptoquark as shown in diagrams in Fig.~\ref{fig:QCDquark}.
\noindent The partonic cross section for this process is given by
\begin{align}
\nonumber\hat{\sigma}^R(z) = \frac{\pi z |y_{q\ell}|^2}{4m_{\rm LQ}^2}\frac{\alpha_s}{2\pi}C_F&\Bigg\{\delta(1-z)\Bigg[-\frac{3}{2}L_\mu^{\rm IR} + \frac{5}{2}\left(\frac{1}{\epsilon_{\rm IR}}+L_\mu^{\rm IR}\right)+\frac{1}{\epsilon_{\rm IR}^2}+\frac{1}{\epsilon_{\rm IR}}L_{\mu}^{\rm IR}
+\frac{1}{2}\left(L_{\mu}^{\rm IR}\right)^2-\frac{\pi^2}{4}+2\Bigg]\\
&+2(1+z^2)\left(\frac{\log(1-z)}{(1-z)}\right)_{+}-\frac{2z}{(1-z)_{+}}-\frac{1+z^2}{(1-z)_{+}}\log\left(\frac{z\mu_{\rm F}^2}{m_{\rm LQ}^2}\right)\Bigg\}.
\end{align}
As expected, the inclusion of the gluon radiation provides the IR divergences that exactly cancel the ones present in the virtual contribution \eqref{eq:virtual}. 
The combined result reads
\begin{gather}
    \hat{\sigma}_{q\ell}(z) = \frac{\pi z |y_{q\ell}|^2}{4 m_{\rm LQ}^2}\Bigg\{\left[1+\frac{\alpha_s}{2\pi}C_F\left(\frac{3}{2}\log\left(\frac{\mu_{\rm R}^2}{\mu_{\rm F}^2}\right)-\frac{\pi^2}{3}\right)\right]\delta(1-z)-\frac{2z}{(1-z)_{+}}\\\nonumber
    +2(1+z^2)\left(\frac{\log(1-z)}{(1-z)}\right)_{+}-\frac{2z}{(1-z)_{+}}-\frac{1+z^2}{(1-z)_{+}}\log\left(\frac{z\mu_{\rm F}^2}{m_{\rm LQ}^2}\right)\Bigg\}\,.
\end{gather}

\begin{figure}[H]
\centering
\includegraphics[width=0.69\textwidth]{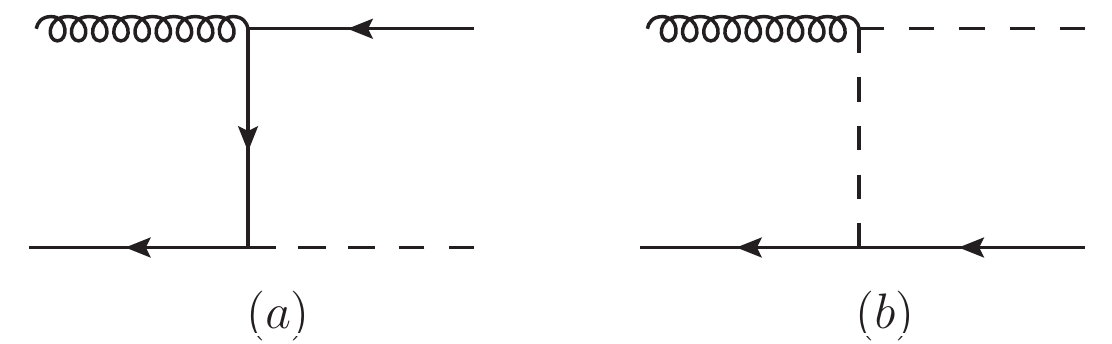}
\caption{Diagrams for the process $g + l\to q + LQ$ contributing to the resonant leptoquark production at $\mathcal{O}(\alpha_s)$.}
\label{fig:QCDglue}
\end{figure}

\noindent The second process which we need to take into account is the one with the soft quark in the final state, corresponding to the diagrams in Fig.~\ref{fig:QCDglue}. 
\noindent The partonic cross section for this process is given by
\begin{equation}
\hat{\sigma}_{g\ell}(z) = \frac{\pi z |y_{q\ell}|^2}{4m_{\rm LQ}^2}\frac{\alpha_s}{2\pi}T_R\left[-\log\left(\frac{z\mu^2_{\rm F}}{(1-z)^2m_{\rm LQ}^2}\right)(z^2+(1-z)^2)+2z(1-z)(2+\log z)\right]\,,
\label{eq:xsecglue}
\end{equation}
where $T_R=1/2$ is the appropriate $SU(3)$ color factor. Due to the massless quark which can be collinear to the gluon in diagram  (a) of Fig. \ref{fig:QCDglue}, this process needs the inclusion of the $\overline{\text{MS}}$ counter-term. This is exactly the same situation we already studied in the process $\gamma + q\to l^- + LQ$, and the universality of the $\log (\mu_{\rm F})$ terms for the two processes in \eqref{eq:xsecglue} and \eqref{eq:gammad}, up to color factors, becomes evident.

%%%%%%%%%%
\section{Supplemental numerical results}
\label{app:supplement}
%%%%%%%%%%

\begin{figure}[tbp]
\centering
\includegraphics[scale=0.65]{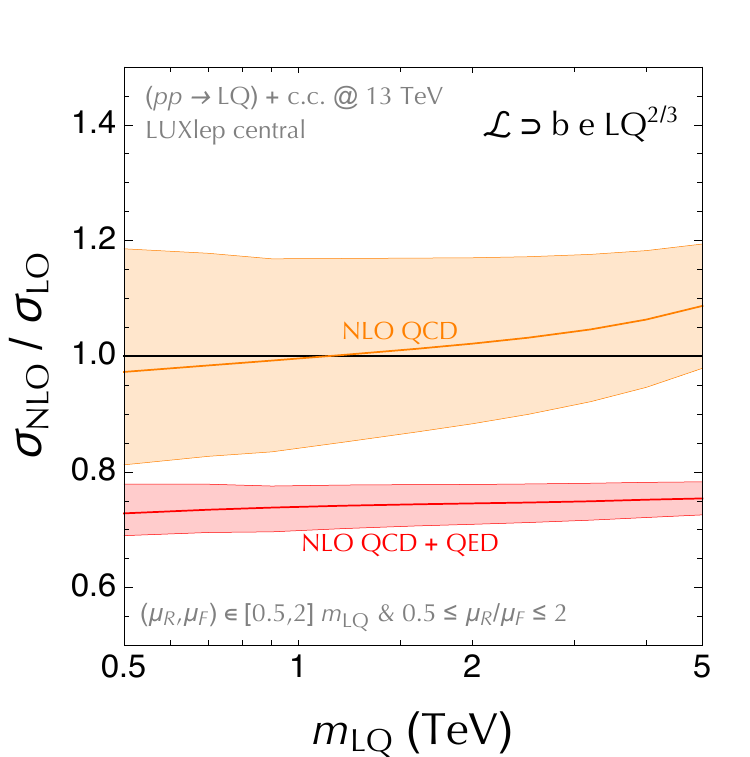} 
\includegraphics[scale=0.65]{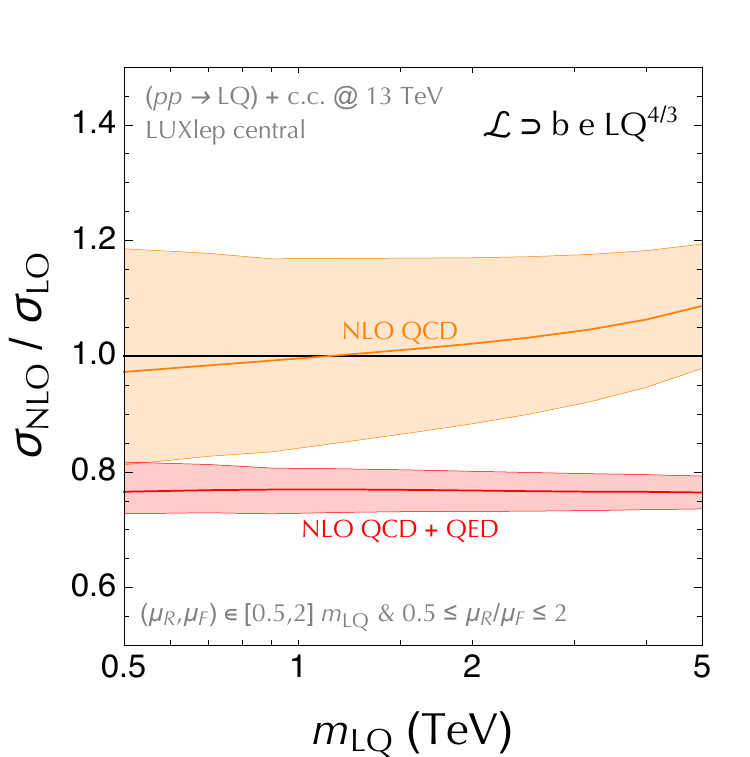} 
\includegraphics[scale=0.65]{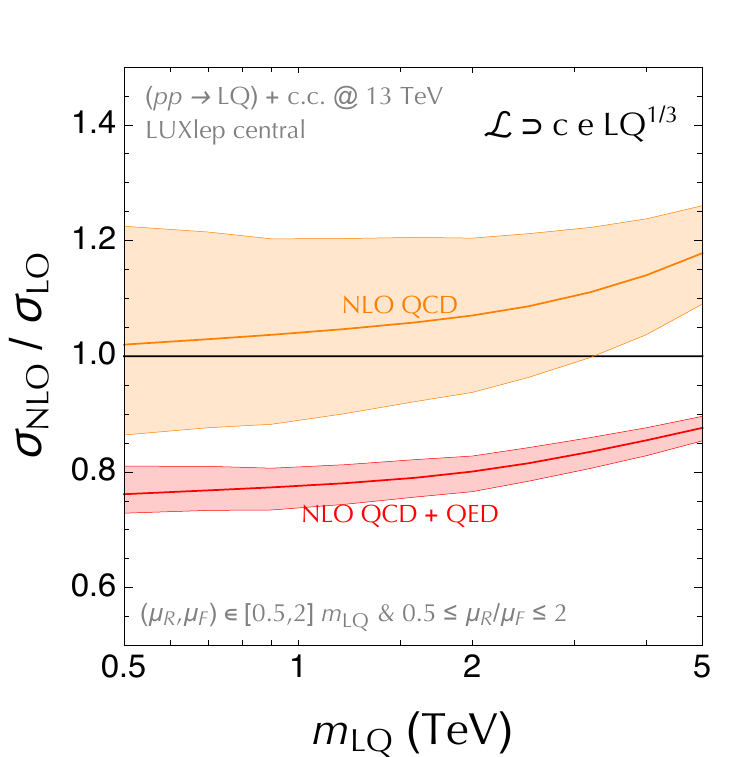}
\includegraphics[scale=0.65]{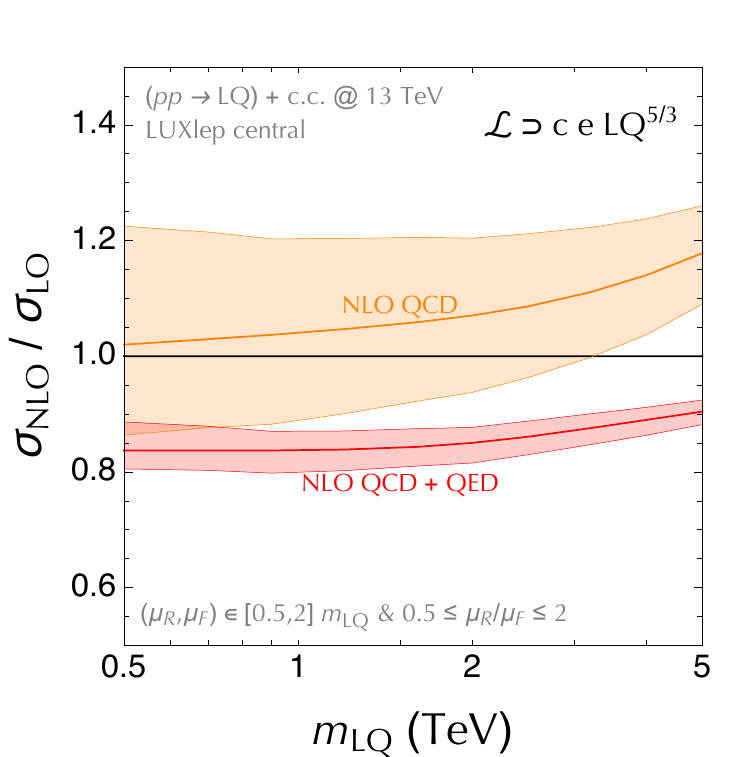} 
\includegraphics[scale=0.65]{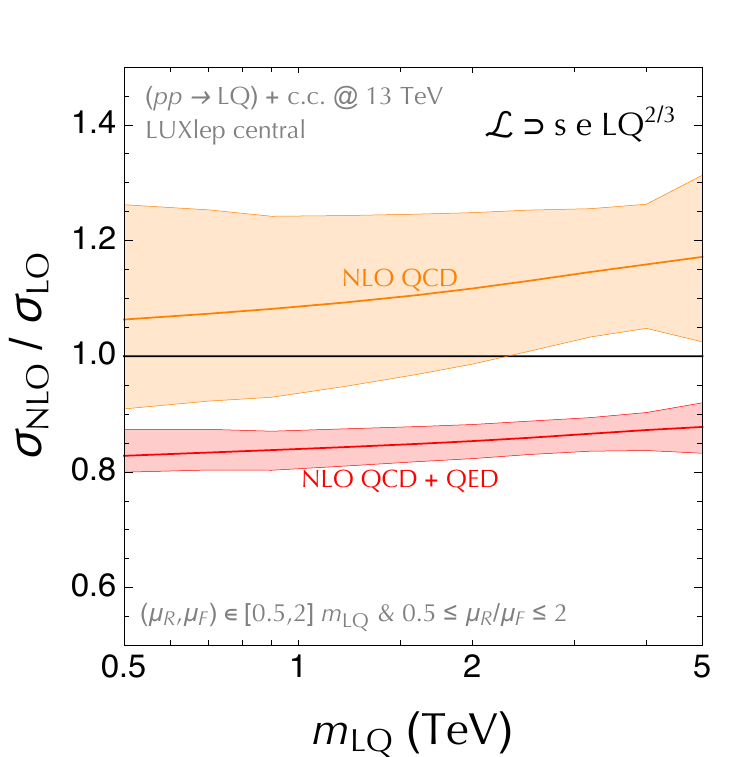} 
\includegraphics[scale=0.65]{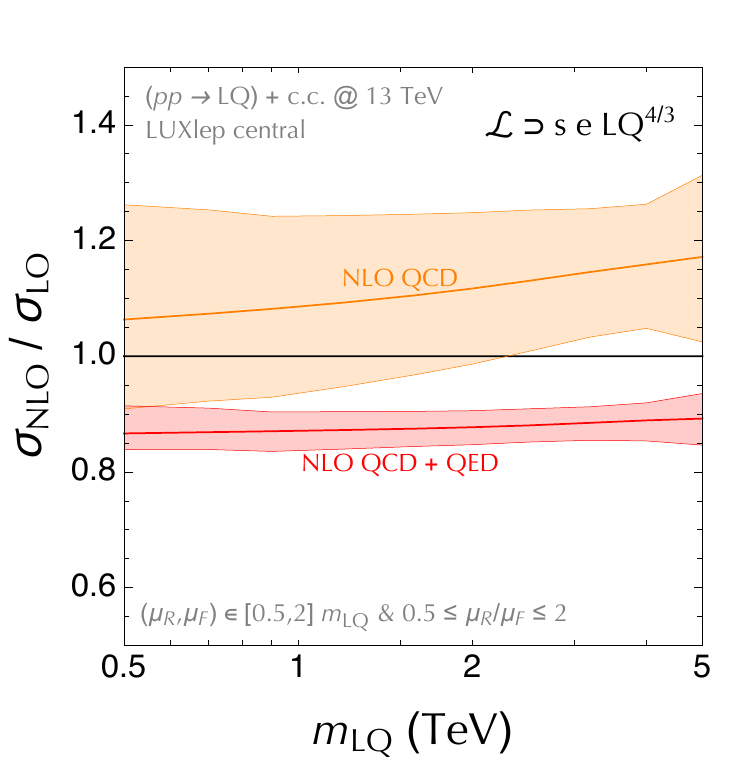}
\includegraphics[scale=0.65]{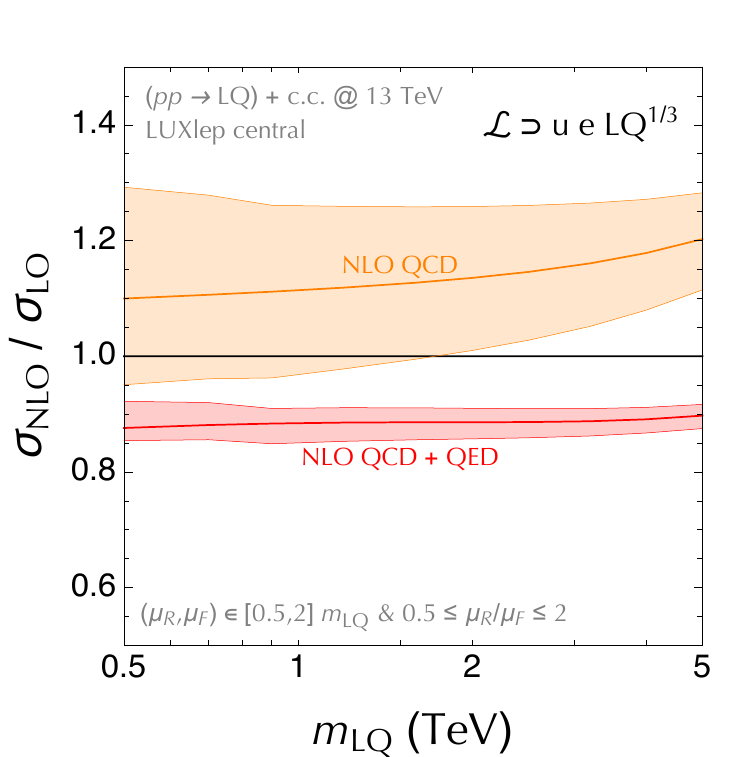} 
\includegraphics[scale=0.65]{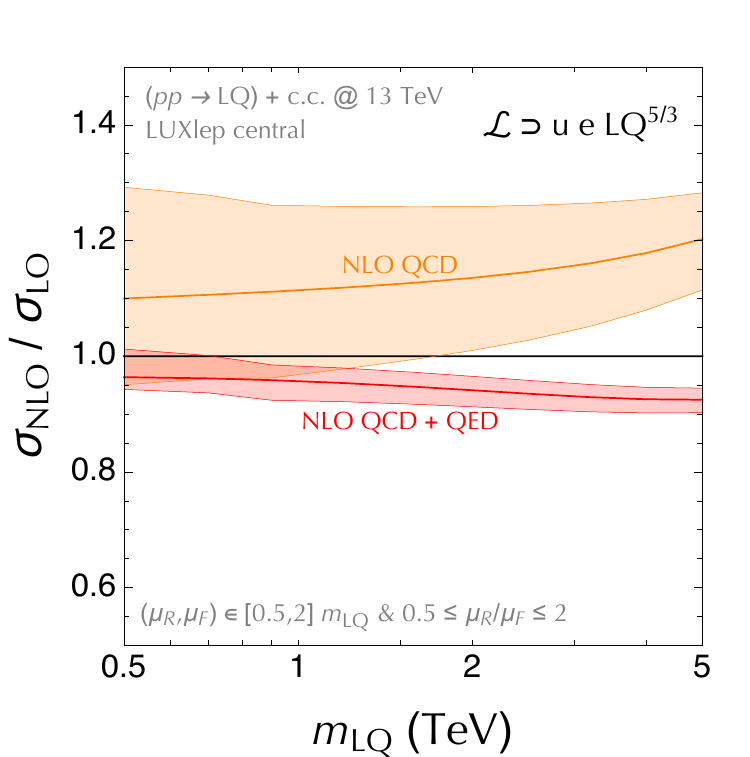} 
\includegraphics[scale=0.65]{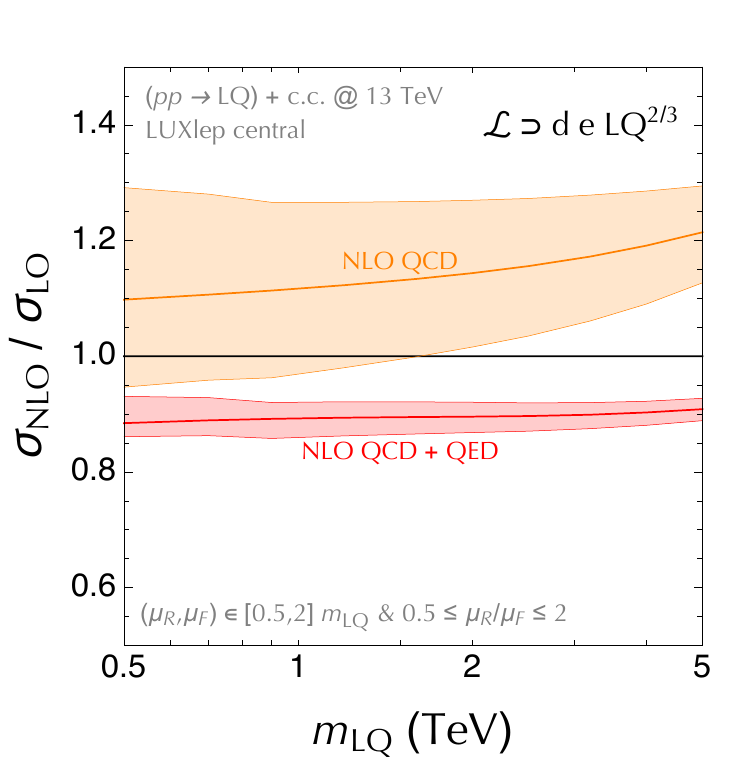}
\includegraphics[scale=0.65]{./Figures/NLOKfactors/512.pdf} 
\caption{\label{fig:SVelectron} ({\bf electron}) NLO $K$-factors ($\sigma_{{\rm NLO}} / \sigma_{{\rm LO}}$) for resonant scalar leptoquark production at 13~TeV LHC. Shown with orange (red) are the NLO QCD (NLO QCD + QED) predictions normalised to the LO when setting the central scales to $\mu_R=\mu_F=m_\mathrm{LQ}$. The colored bands are obtained by varying factorisation and renormalisation scales in the NLO calculations within $\{ \mu_F, \mu_R \} \in [0.5, 2]~ m_\mathrm{LQ}$ while respecting $1/2 \le \mu_R/\mu_F \le 2$. We use the central PDF set from {\tt LUXlep-NNPDF31\_nlo\_as\_0118\_luxqed} (v2)~\cite{Buonocore:2020nai}. }
\end{figure}

\begin{figure}[tbp]
\centering
\includegraphics[scale=0.65]{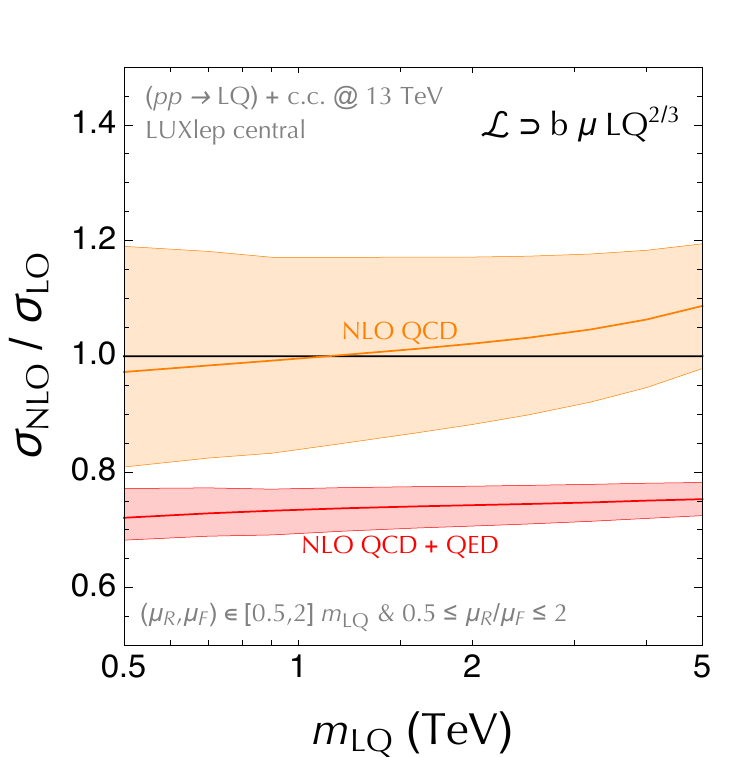} 
\includegraphics[scale=0.65]{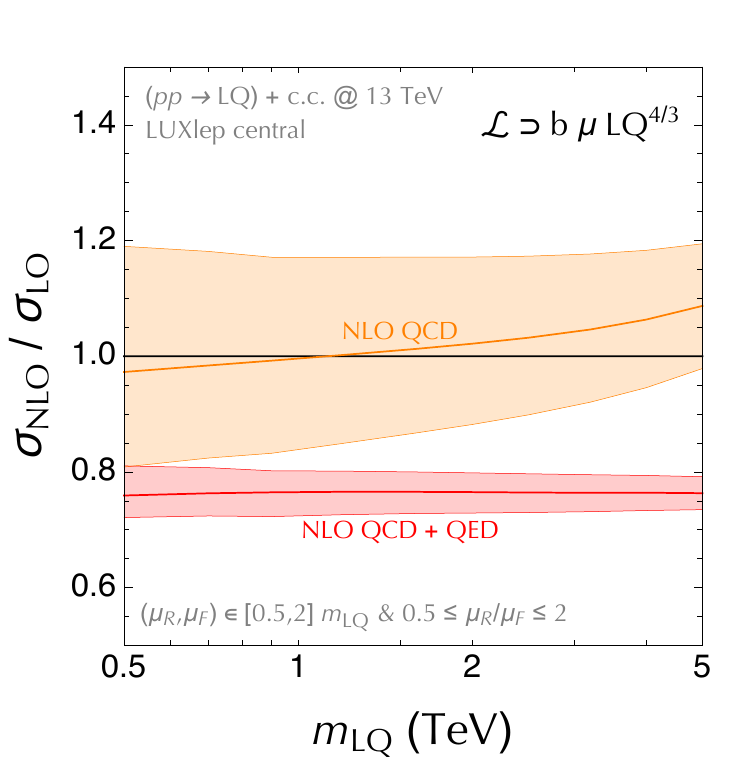} 
\includegraphics[scale=0.65]{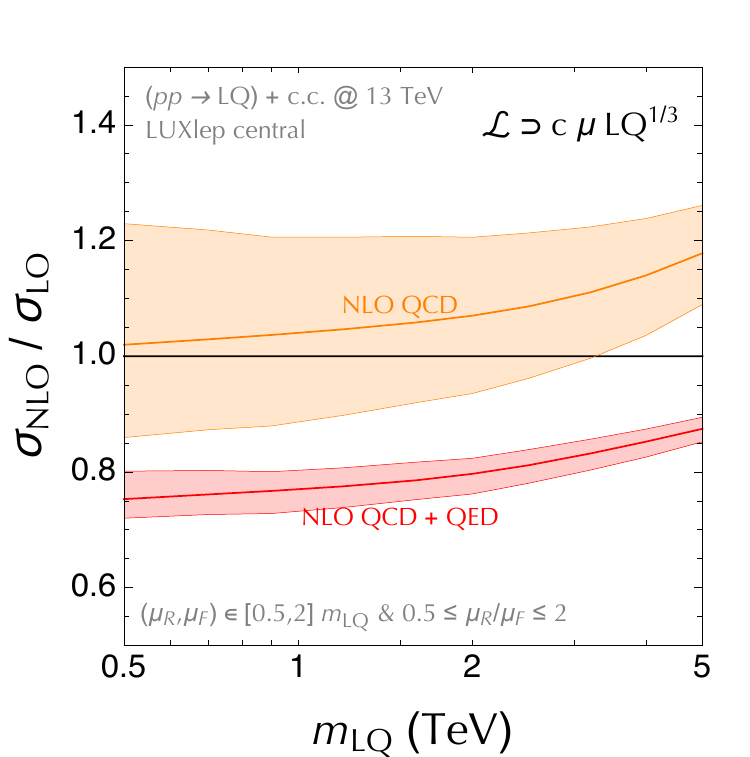}
\includegraphics[scale=0.65]{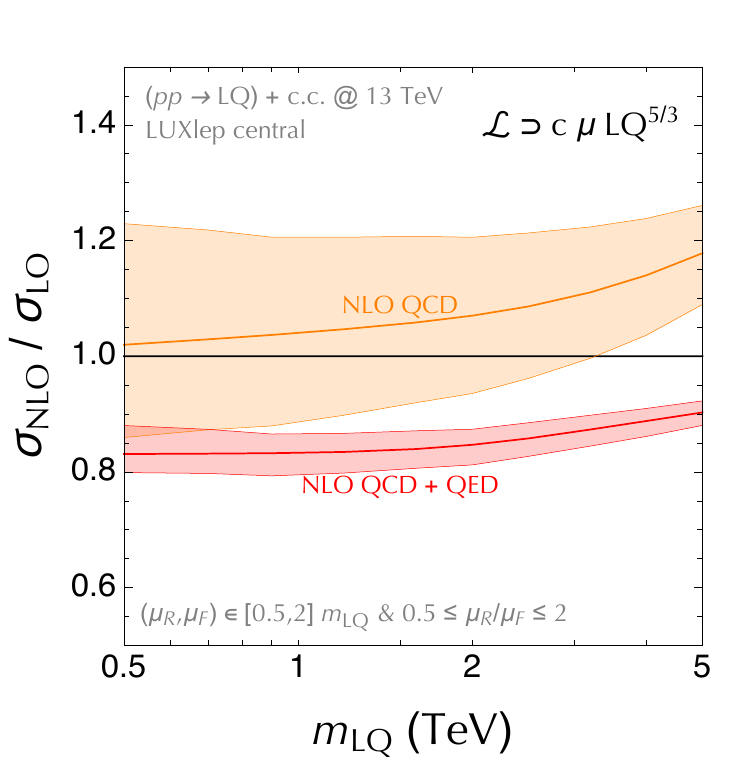} 
\includegraphics[scale=0.65]{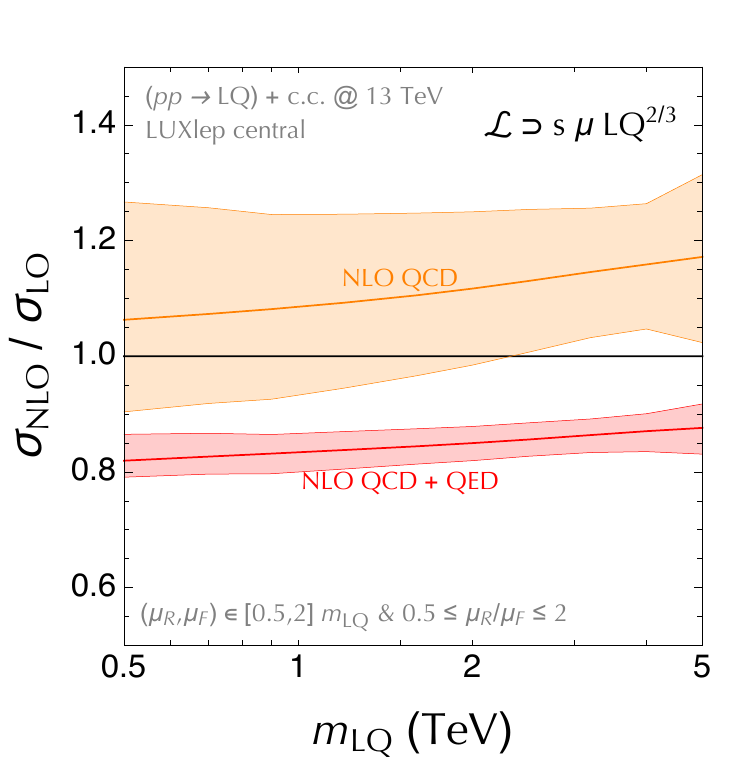} 
\includegraphics[scale=0.65]{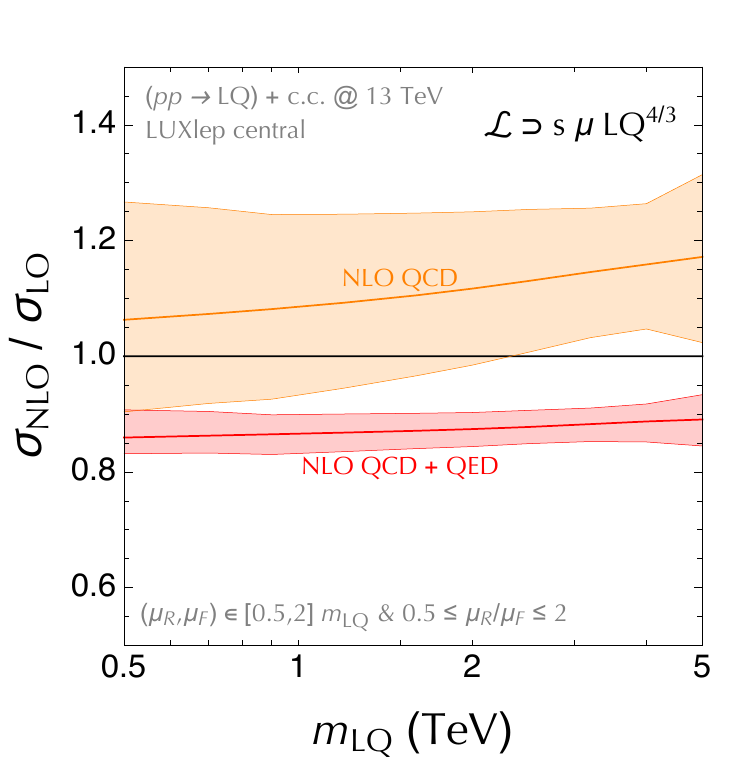}
\includegraphics[scale=0.65]{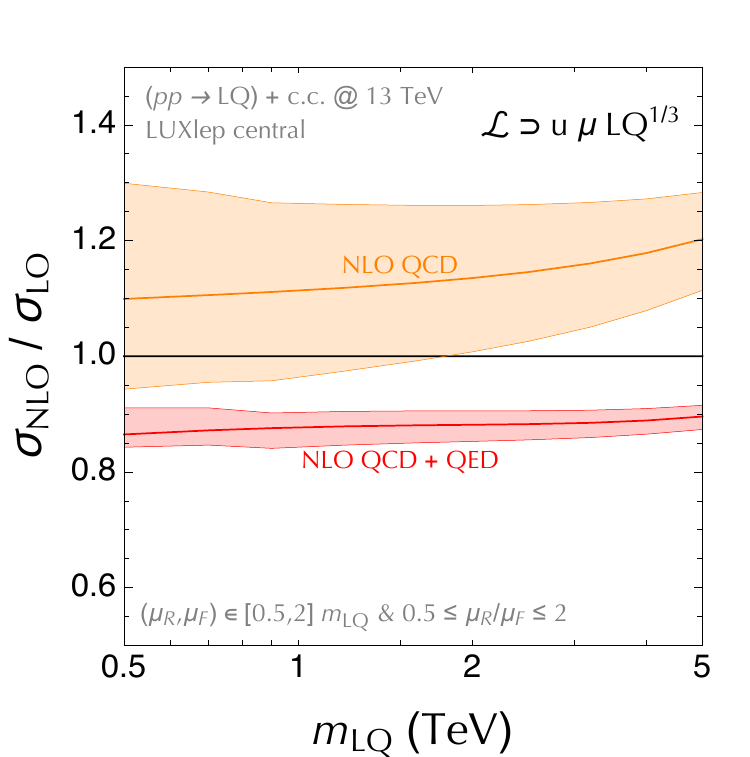} 
\includegraphics[scale=0.65]{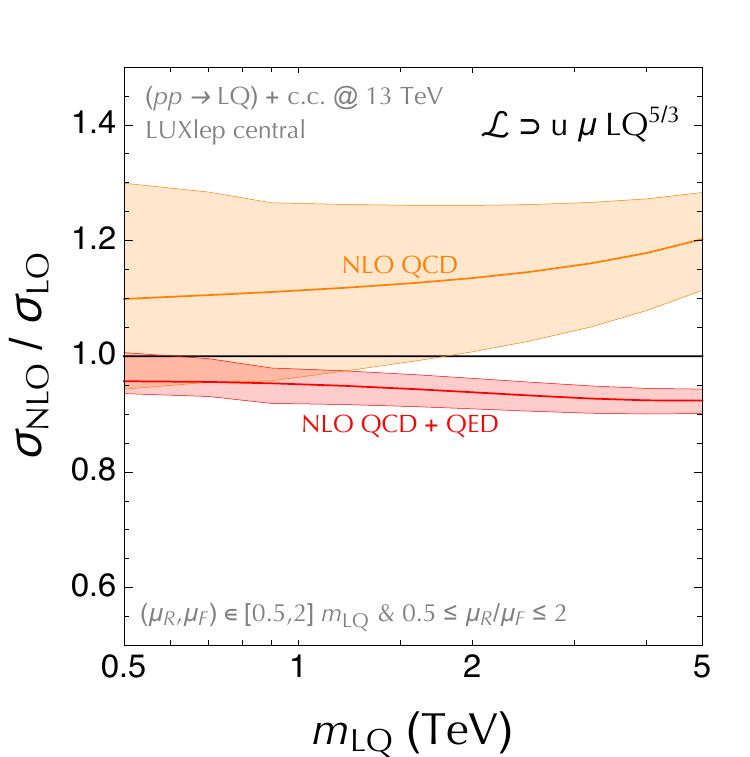} 
\includegraphics[scale=0.65]{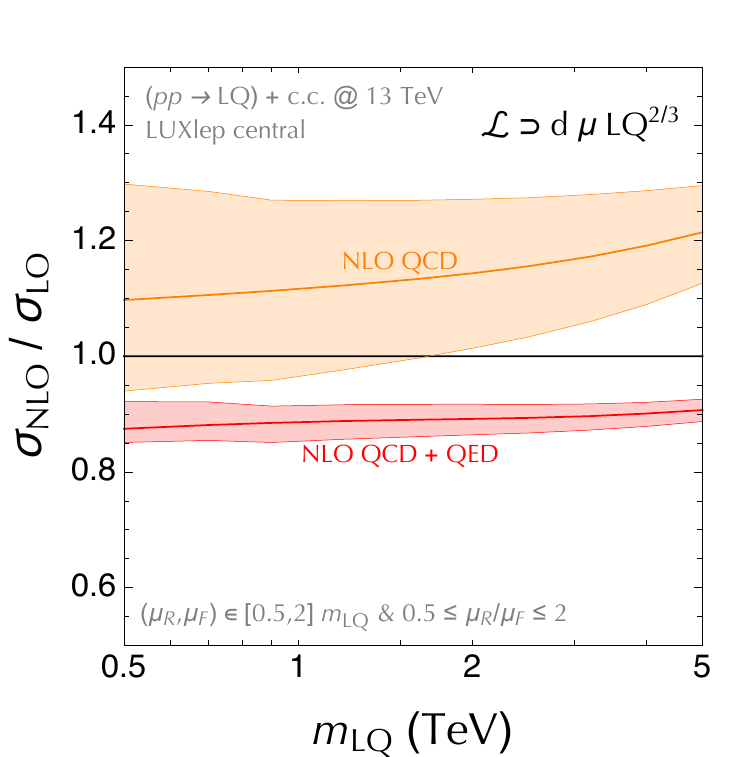}
\includegraphics[scale=0.65]{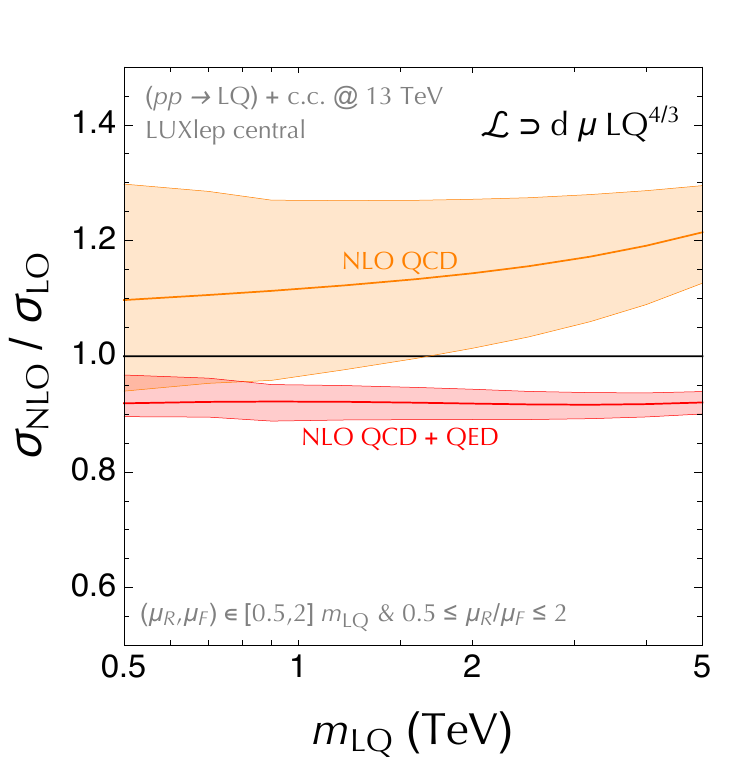} 
\caption{\label{fig:SVmuon} ({\bf muon}) NLO $K$-factors ($\sigma_{{\rm NLO}} / \sigma_{{\rm LO}}$) for resonant scalar leptoquark production at 13~TeV LHC. Shown with orange (red) are the NLO QCD (NLO QCD + QED) predictions normalised to the LO when setting the central scales to $\mu_R=\mu_F=m_\mathrm{LQ}$. The colored bands are obtained by varying factorisation and renormalisation scales in the NLO calculations within $\{ \mu_F, \mu_R \} \in [0.5, 2]~ m_\mathrm{LQ}$ while respecting $1/2 \le \mu_R/\mu_F \le 2$. We use the central PDF set from {\tt LUXlep-NNPDF31\_nlo\_as\_0118\_luxqed} (v2)~\cite{Buonocore:2020nai}. }
\end{figure}

\begin{figure}[tbp]
\centering
\includegraphics[scale=0.65]{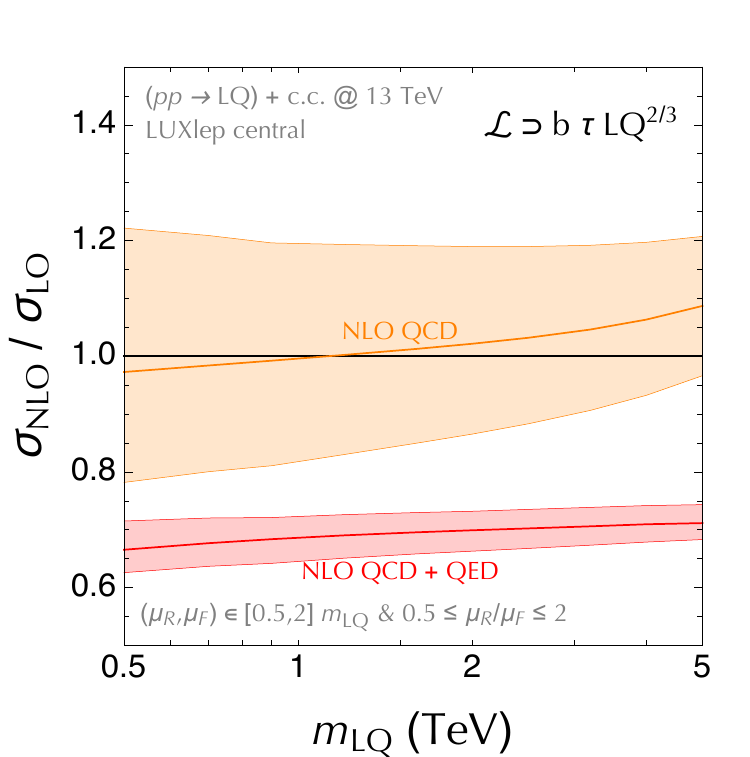} 
\includegraphics[scale=0.65]{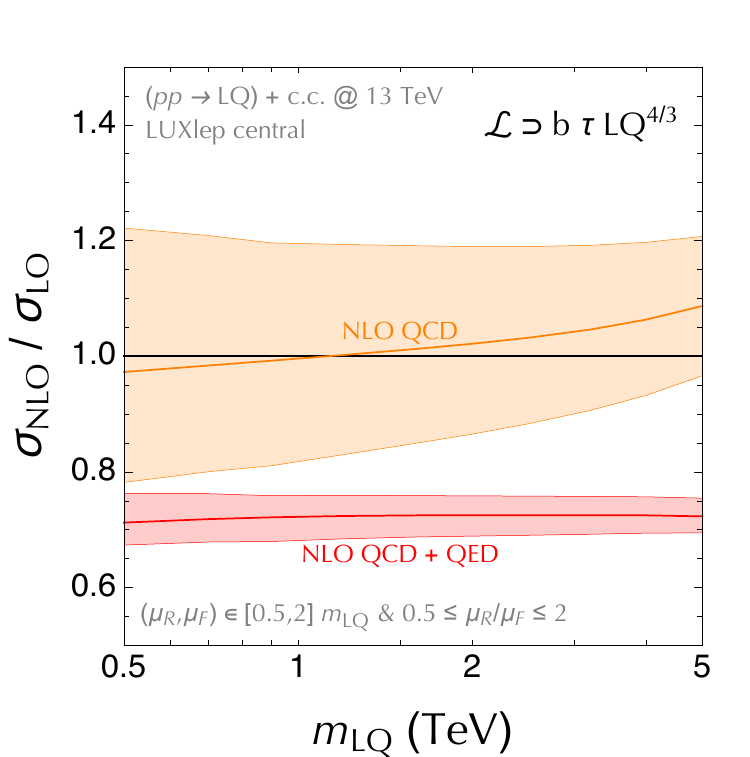} 
\includegraphics[scale=0.65]{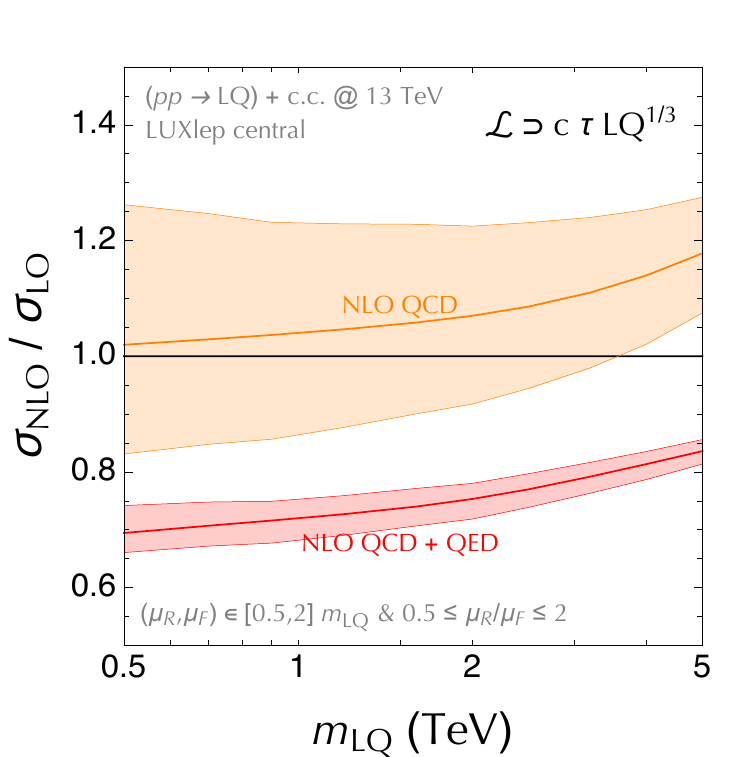}
\includegraphics[scale=0.65]{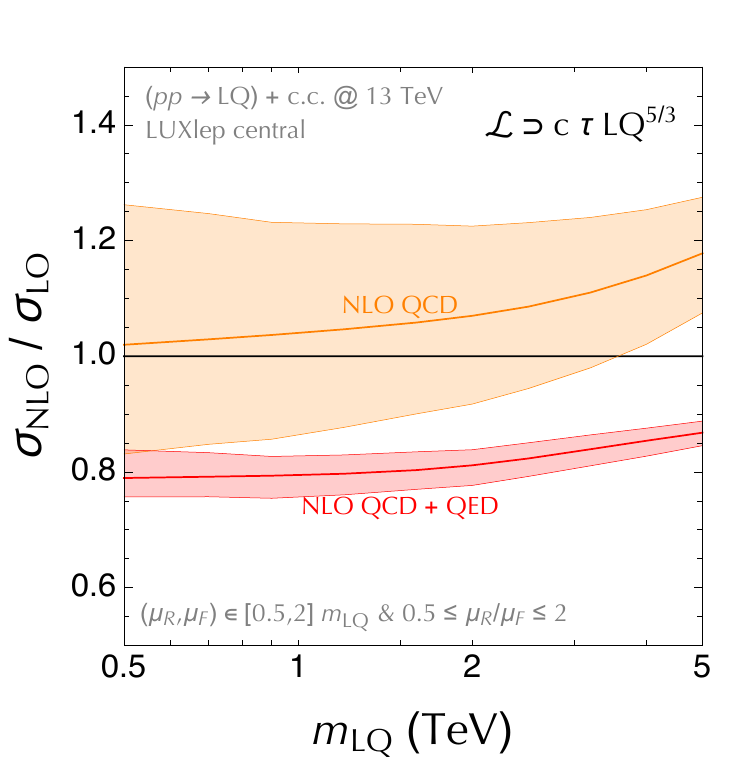} 
\includegraphics[scale=0.65]{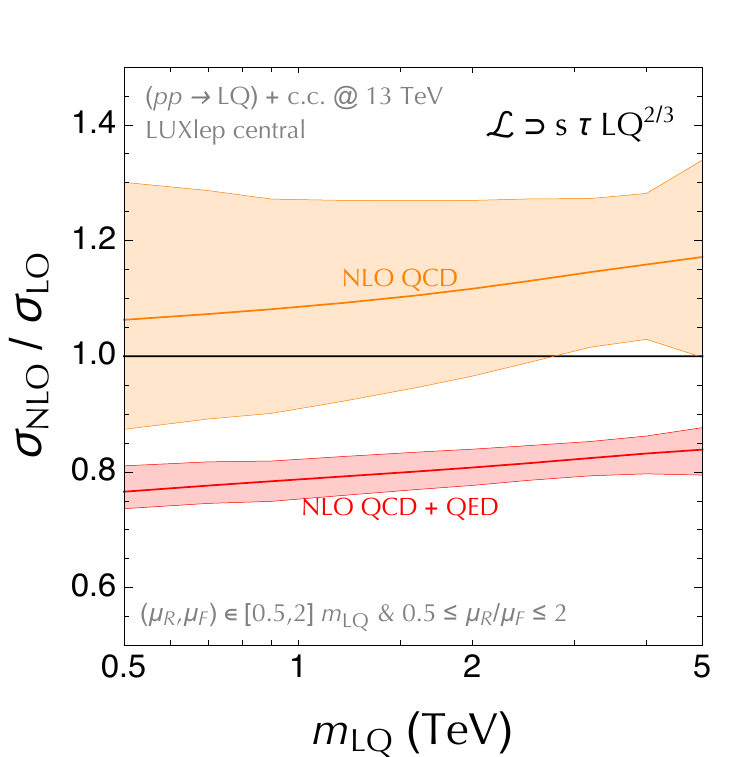} 
\includegraphics[scale=0.65]{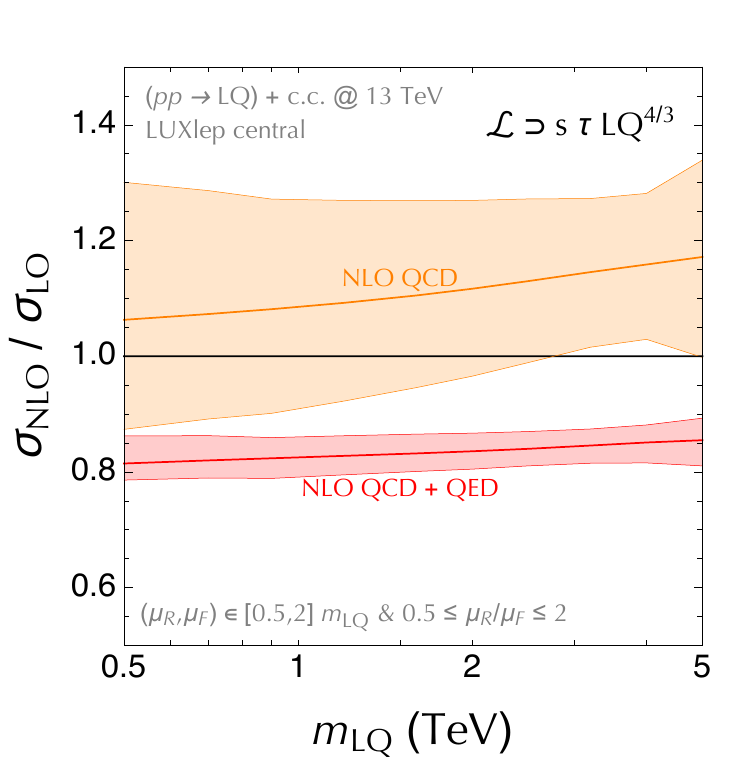}
\includegraphics[scale=0.65]{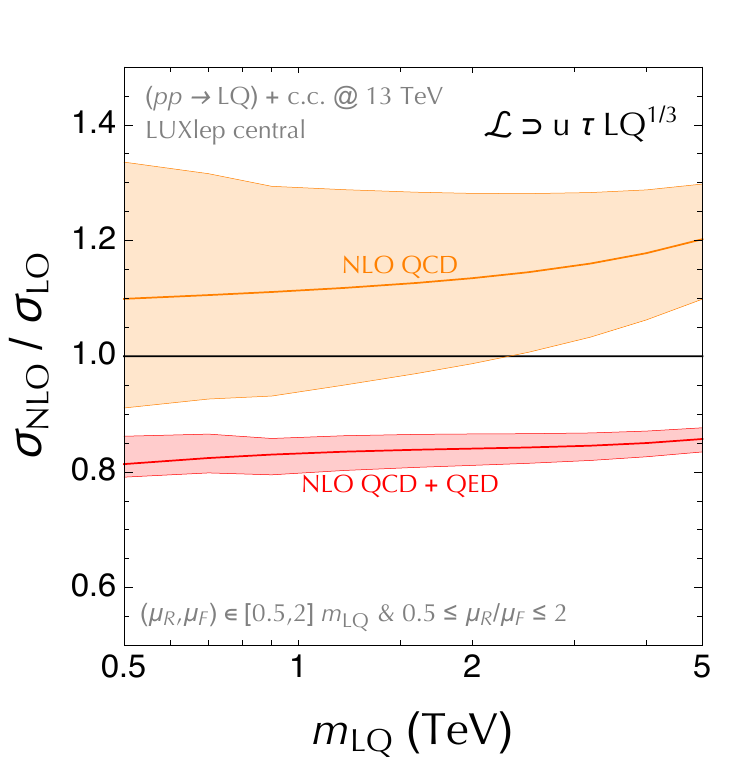} 
\includegraphics[scale=0.65]{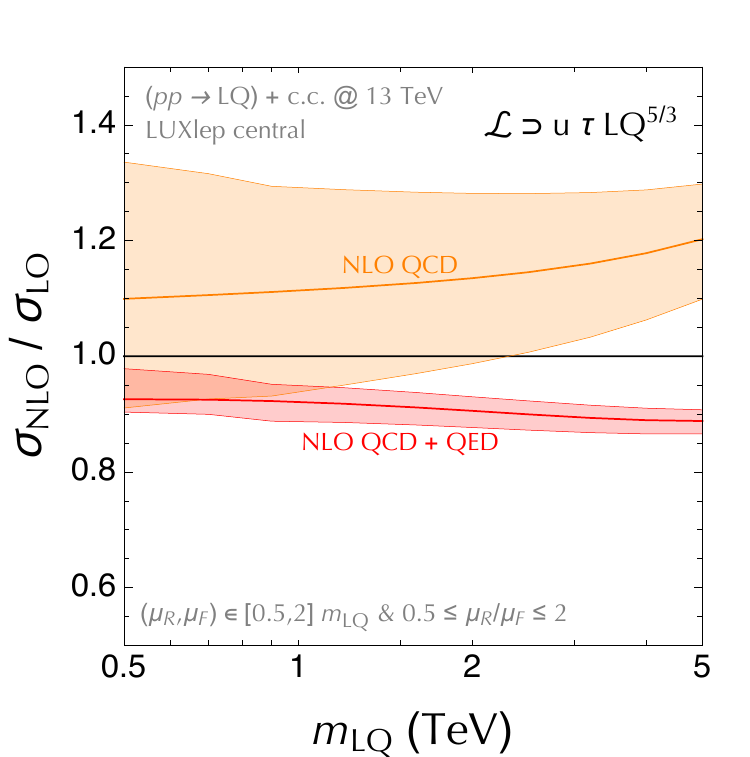} 
\includegraphics[scale=0.65]{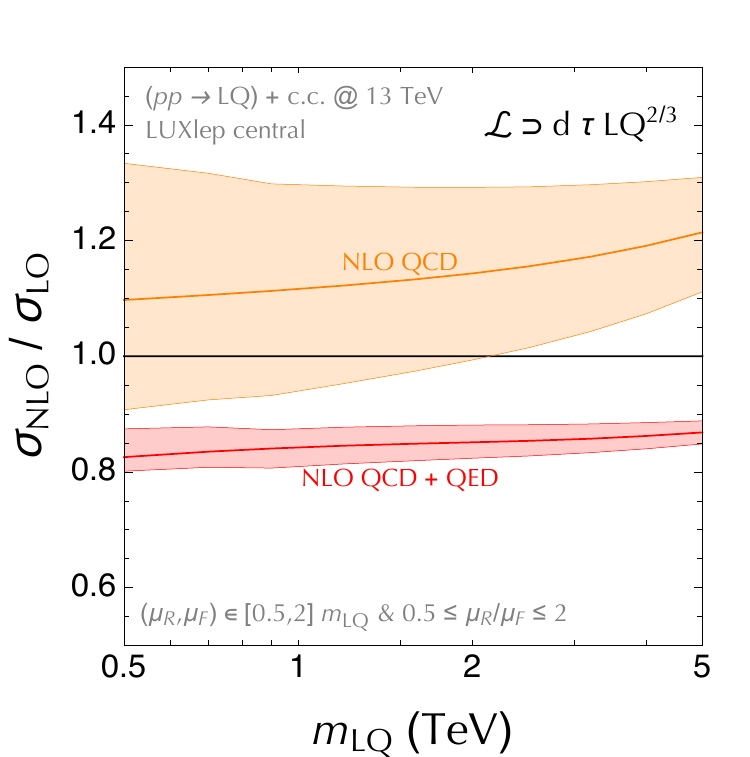}
\includegraphics[scale=0.65]{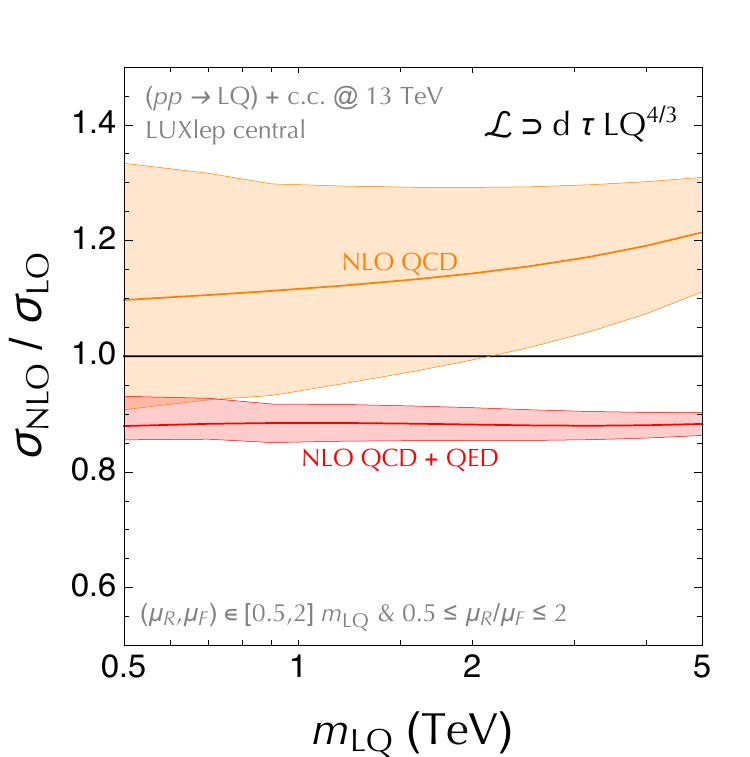} 
\caption{\label{fig:SVtau} ({\bf tau}) NLO $K$-factors ($\sigma_{{\rm NLO}} / \sigma_{{\rm LO}}$) for resonant scalar leptoquark production at 13~TeV LHC. Shown with orange (red) are the NLO QCD (NLO QCD + QED) predictions normalised to the LO when setting the central scales to $\mu_R=\mu_F=m_\mathrm{LQ}$. The colored bands are obtained by varying factorisation and renormalisation scales in the NLO calculations within $\{ \mu_F, \mu_R \} \in [0.5, 2]~ m_\mathrm{LQ}$ while respecting $1/2 \le \mu_R/\mu_F \le 2$. We use the central PDF set from {\tt LUXlep-NNPDF31\_nlo\_as\_0118\_luxqed} (v2)~\cite{Buonocore:2020nai}. }
\end{figure}

Figs.~\ref{fig:SVelectron},~\ref{fig:SVmuon}, and~\ref{fig:SVtau} show the NLO K-factors for three lepton flavors: electron, muon and tau, respectively. Each figure contains ten plots for different quark flavors and leptoquark charge. Down-type quarks couple the $|Q_{{\rm LQ}}|=2/3, 4/3$, while the up-type quarks couple the $|Q_{{\rm LQ}}|=1/3, 5/3$. For more details see Section~\ref{sec:dis}.

\bibliographystyle{JHEP}
\bibliography{references}
	
\end{document}